\documentclass{aa}  

\usepackage[colorlinks, allcolors=blue]{hyperref}
\usepackage{amsmath} 

\usepackage{graphicx}
\usepackage{xcolor}
\usepackage{txfonts}
\usepackage{longtable}
\usepackage{geometry}
\usepackage{datetime}
\usepackage{lscape}
\usepackage{float}
\usepackage{url}
\usepackage{orcidlink}

\def\arcsec{$^{\prime\prime}$}

\begin{document} 

\title{Probing multi-band variability and mode switching in the candidate transitional millisecond pulsar 3FGL J1544.6$-$1125}

\author{G.~Illiano\inst{1,2} \orcidlink{0000-0003-4795-7072}, 
 F.~Coti Zelati\inst{3,4,1} \orcidlink{0000-0001-7611-1581}, 
 A.~Miraval Zanon\inst{5} \orcidlink{0000-0002-0943-4484}, 
 A.~Papitto\inst{2} \orcidlink{0000-0001-6289-7413}, 
 M.~C.~Baglio\inst{1} \orcidlink{0000-0003-1285-4057}, 
 D.~de Martino\inst{6} \orcidlink{0000-0002-5069-4202},
 S.~Giarratana \inst{1} \orcidlink{0000-0002-2815-7291},
 F.~Ambrosino\inst{2} \orcidlink{0000-0001-7915-996X},
 F.~Carotenuto\inst{2} \orcidlink{0000-0002-0426-3276},
 S.~Campana\inst{1} \orcidlink{0000-0001-6278-1576},
 A.~Marino\inst{3,4,7} \orcidlink{0000-0001-5674-4664},
 N.~Rea\inst{3,4} \orcidlink{0000-0003-2177-6388},
 D.~F.~Torres\inst{3,8,4} \orcidlink{0000-0002-1522-9065},
 M.~Giroletti\inst{9} \orcidlink{0000-0002-8657-8852},
 T.~D.~Russell\inst{7} \orcidlink{0000-0002-7930-2276},
 C.~Malacaria\inst{2} \orcidlink{0000-0002-0380-0041}, 
 C.~Ballocco\inst{2} \orcidlink{0009-0001-0155-7455}, 
 E.~Bozzo\inst{10} \orcidlink{0000-0002-8201-1525},
 C.~Ferrigno\inst{10} \orcidlink{0000-0003-1429-1059},
 R.~La Placa\inst{2} \orcidlink{0000-0003-2810-2394}, 
 A.~Ghedina\inst{11} \orcidlink{0000-0003-4702-5152},
 M.~Cecconi \inst{11},
 F.~Leone\inst{12,13} \orcidlink{0000-0001-7626-3788}
 }

 \institute{
 INAF–Osservatorio Astronomico di Brera, Via Bianchi 46, I-23807, Merate (LC), Italy
 \and
 INAF-Osservatorio Astronomico di Roma, Via Frascati 33, I-00078, Monte Porzio Catone (RM), Italy\\
 \email{giulia.illiano@inaf.it}
 \and Institute of Space Sciences (ICE, CSIC), Campus UAB, Carrer de Can Magrans s/n, E-08193 Barcelona, Spain
 \and Institut d’Estudis Espacials de Catalunya (IEEC), E-08860 Castelldefels (Barcelona), Spain  
 \and ASI - Agenzia Spaziale Italiana, Via del Politecnico snc, 00133 Roma, Italy
 \and INAF–Osservatorio Astronomico di Capodimonte, Salita Moiariello 16, I-80131 Naples, Italy
 \and INAF, Istituto di Astrofisica Spaziale e Fisica Cosmica, Via U. La Malfa 153, I-90146 Palermo, Italy
 \and Instituci\'o Catalana de Recerca i Estudis Avan\,cats (ICREA), E-08010 Barcelona, Spain 
 \and INAF – Istituto di Radioastronomia, Via Gobetti 101, 40129, Bologna, Italy
 \and Department of Astronomy, University of Geneva, chemin d’Écogia, 16, 1290 Versoix, Switzerland
 \and Fundación Galileo Galilei – INAF, Rambla J.A.Fernández P. 7, 38712, B.Baja, (S.C.Tenerife), Spain
 \and Dipartimento di Fisica e Astronomia, Sezione Astrofisica, Università di Catania, Via S. Sofia 78, I-95123 Catania, Italy
 \and INAF–Osservatorio Astrofisico di Catania, Via S. Sofia 78,I-95123 Catania, Italy
 }             

\date{}
\authorrunning{Illiano et al.}
\titlerunning{Multi-band variability and mode switching in the candidate tMSP 3FGL~J1544.6$-$1125}
 
\abstract{
We present the most extensive high-time resolution multi-band campaign to date on the candidate transitional millisecond pulsar (tMSP) 3FGL~J1544.6$-$1125 in the enigmatic sub-luminous disk state, with coordinated observations from the radio to the X-ray band. 
While \textit{XMM‑Newton} and \textit{NuSTAR} X‑ray light curves exhibit the characteristic high‑ and low‑mode bimodality, the source faintness prevents firm evidence for similar bimodality in the ultraviolet and near‑infrared light curves, presented here for the first time. A re-analysis of archival \textit{XMM‑Newton}/OM data reveals an optical flare without an X‑ray counterpart, likely originating from the outer accretion disk or the companion star.
During our observations, no radio emission was detected, with a 3$\sigma$ flux density upper limit of $\sim$8\,$\mu$Jy at 6\,GHz. While past works have already reported radio variability in the source, this limit is a factor of $\gtrsim$3.5 below the average value measured in 2019 under similar conditions, underscoring significant radio variability despite the relatively stable X‑ray flux.
Simultaneous optical light curves in five filters obtained with GTC/HiPERCAM revealed flickering and dipping activities that resemble the observed X-ray variability, along with a reddening trend at lower fluxes. The latter is consistent with discrete mass ejections that disrupt the inner flow and reduce both X-ray and optical fluxes, thereby driving the high-to-low-mode switches.
This suggests a common origin for most optical and X‑ray emission at the boundary region between the pulsar wind and the inner disk, as also supported by our modelling of the spectral energy distribution in the high mode.
Overall, our findings reinforce the mini-pulsar nebula picture for tMSPs in the sub-luminous state and demonstrate how coordinated, high-time-resolution, multi-wavelength campaigns are essential to understand the physical processes governing rapid mode switches in these systems.
}


\maketitle

\section{Introduction} \label{sec:intro}
Millisecond pulsars (MSPs) are the fastest-spinning neutron stars (NSs) in the Universe. They achieve their millisecond spin period during an accretion phase lasting billions of years. Throughout this time, a low-mass ($\lesssim 1 \, \mathrm{M_\odot}$) donor star transfers matter and angular momentum to the NS surface, causing the system to shine as a bright low-mass X-ray binary (LMXB). Once the mass accretion rate ceases, the NS transitions to a radio pulsar state driven by the gradual loss of its rotational energy \citep{Alpar_1982Natur, Radhakrishnan_Srinivasan_1982CSci}. MSPs found in tight (i.e., with an orbital period $P_\mathrm{orb} < 1 \, \mathrm{d}$) binary systems provide crucial clues for investigating the complex mechanisms behind mass accretion and ejection as well as the processes of magnetospheric particle acceleration.

Among MSPs, an intriguing sub-class is represented by transitional ones (tMSPs; \citealt[][and references therein]{Papitto_deMartino_2022ASSL}). These unique objects have conclusively bridged the long-sought evolutionary gap between accreting NSs and radio MSPs in binary systems. Remarkably, tMSPs can swing back and forth between accretion-powered and rotation-powered states on timescales accessible within a human lifetime, from years to decades. All confirmed tMSPs and a few candidates have also been caught in a third state, which, based on its characteristics, appears to be intermediate between the two standard states discussed above. This is referred to as the sub-luminous disk state. This state is characterized by the presence of an accretion disk, an X-ray luminosity that is fainter than in the fully accreting state but brighter than in the rotation-powered phase, and a $\gamma$-ray luminosity up to ten times higher than that of MSPs in their state fed by rotation \citep[see, e.g.,][and references therein]{Torres_Li_2022ASSL}. The most defining feature is the variable X-ray emission that unpredictably switches between two distinct intensity levels, dubbed `high' and `low' modes \citep{Papitto_2013Natur, Linares_2014}, along with occasional flares \citep[e.g.,][]{Tendulkar_2014, Bogdanov_2015ApJ_J1023}. 
For confirmed tMSPs, the high mode is dominant, typically occurring for 65–80\% of the time \citep[e.g.,][]{Archibald_2015, Bogdanov_2015ApJ_J1023, deMartino_2013A&A, Papitto_2013Natur, Linares_2014}. A similar behavior has been observed in the candidate tMSP CXOU~J110926.4$-$650224 \citep{CotiZelati2024}. In contrast, some other candidates spend a smaller fraction of time in the high mode, such as around 45–50\% for 3FGL\,J1544.6$-$1125 (hereafter J1544; \citealt{Gusinskaia_2025MNRAS}) and below $40\%$ for 4FGL~J0407.7$-$5702 \citep{Miller_2020}, while 3FGL~J0427.9$-$6704 seems to spend most of its time in an X-ray flaring mode \citep{Li_2020ApJ}. However, continued long‐term monitoring is needed to better establish the long-term behavior of these systems.

\subsection{The archetype transitional PSR~J1023$+$0038}
The archetype of tMSPs, PSR~J1023$+$0038 (hereafter J1023; \citealt{Archibald_2009Sci}) has been continuously observed in all energy ranges since its turn-on from a long rotation-powered state. Its X-ray emission switches from high to low mode on a time scale of $\lesssim$10\,s, whereas the switch from low to high mode takes $\approx$30\,s \citep{Baglio_CotiZelati_2023A&A}.
Flickering and flaring activities are also observed across UV, optical, and near-infrared bands. 
UV mode-switching, correlating with X-rays, was clearly identified through high-time resolution observations performed with the \textit{Hubble} Space Telescope \citep{Jaodand2021, Miraval_Zanon_2022, Baglio_CotiZelati_2023A&A}.
High-cadence optical observations \citep{Shahbaz_2015, Shahbaz_2018, Hakala2018} revealed sharp-edged, rectangular, flat-bottomed dips, strongly resembling X-ray bimodal behavior \citep[see also][]{Kennedy_2018, Papitto_2018}. On the other hand, evidence for near-infrared bimodal variability is limited. Dips in this band appear indeed less pronounced than those at higher energies, possibly due to the enhanced contribution from the outer disk regions and the irradiated companion star \citep{Baglio2019, Papitto_2019ApJ, Baglio_CotiZelati_2023A&A}. Simultaneous radio and X-ray observations showed an anti-correlated variability pattern \citep{Bogdanov2018}. Specifically, when the source switched from the high to the low X-ray mode, a brightening in radio emission was observed, and vice versa. This behavior has been interpreted as rapid ejections of plasmoids during high-to-low modes switches (\citealt{Bogdanov2018}; see also \citealt{Baglio_CotiZelati_2023A&A}). 
X-ray, UV, and optical pulsations at the NS spin period have been detected exclusively during the high mode \citep{Archibald_2015, Ambrosino_2017, Papitto_2019ApJ, Jaodand2021, Miraval_Zanon_2022}. This, along with the similar pulse shapes and the fact that the spectral energy distribution (SED) of the pulsed emission from optical to X-ray wavelengths follows a single power-law relation \citep{Papitto_2019ApJ, Miraval_Zanon_2022}, suggests a common underlying mechanism for the optical/UV and X-ray pulsations.
Recently, it was proposed the so-called ``mini-pulsar nebula scenario'', suggesting that during the high mode, X-ray/UV/optical pulsations originate from synchrotron radiation in the boundary region where the particle wind ejected from a rotation-powered pulsar meets the matter from the inner accretion disk at about 100\,km from the pulsar (\citealt{Papitto_2019ApJ, Veledina_2019}; see also \citealt{Illiano_2023AA}). Recent \textit{IXPE} polarimetric observations revealed polarized X-ray emission with a polarization degree up to 15\% during the high mode, strongly supporting the rotation-powered pulsar wind model \citep{Baglio2024}.
The switch to the low mode is believed to be triggered by discrete mass ejections \citep{Baglio_CotiZelati_2023A&A}, which remove material from the inner disk and cause a decline in the X-ray, UV, and optical flux. As the flow from the disk refills the inner regions, the boundary region front is restored, and the system switches back to the high mode. 

\subsection{The candidate transitional 3FGL\,J1544.6$-$1125}
J1544, initially identified as a $\gamma$-ray source in the third \textit{Fermi}-LAT source catalogue \citep{Acero_2015ApJS}, has emerged as a robust candidate tMSP in the sub-luminous disk state due to its strikingly similar X-ray and $\gamma$-ray emission properties to those of J1023 \citep{Bogdanov_2015ApJ}. 
Specifically, the X-ray emission of J1544 exhibits a clear bimodal behavior, alternating between high- and low-intensity modes \citep{Bogdanov_2015ApJ, Bogdanov_2016ApJ}, and shows a clear dipping behavior in the optical band \citep{Bogdanov_2015ApJ}. The spectrum of the X-ray emission is well described by an absorbed power law with a photon index of $\Gamma \sim 1.7$ \citep{Bogdanov_2016ApJ}, consistent with typical spectra observed for tMSPs during their intermediate states \citep[e.g.,][]{Papitto_deMartino_2022ASSL}. The broadband SED from the optical band to the $\gamma$-rays closely resembles that of J1023 \citep[Fig 4. from][]{Bogdanov_2016ApJ}. Further evidence for classifying J1544 as a tMSP comes from its position on the radio/X-ray luminosity plane (for an X-ray luminosity of $5.51 \times 10^{33} \, \mathrm{erg \, s^{-1}}$; \citealt{Jaodand_2021ApJ}), which is very close to the region occupied by the confirmed tMSPs J1023 and XSS~J12270$-$4859. Additionally, J1544 shows variable radio continuum emission on timescales of a few days, akin to the behavior observed in J1023 \citep{Jaodand_2021ApJ, Gusinskaia_2025MNRAS}. Hints of an anti-correlated variability pattern between simultaneous radio and X-ray observations have been reported by \citet{Gusinskaia_2025MNRAS}, resembling that seen in J1023.
Optical spectroscopy revealed that J1544 is nearly face-on and in a $\sim$5.8 hr orbit around a likely mid-to-late K-type main-sequence star \citep{Britt_2017ApJ}.
Using graph theory algorithms, \citet{Garcia_2024A&A} recently showed that, given the system parameters and their associated uncertainties, J1544 predominantly clusters with confirmed tMSPs, thereby reinforcing its classification as a highly promising candidate.

To ultimately validate J1544 as a tMSP, it is necessary to observe a transition either to a rotation-powered or accretion-powered state with detectable millisecond pulsations. At the time of writing, IGR~J18245$-$2452 is the only tMSP to have exhibited a bright accretion outburst during which 3.9-ms X-ray pulsations were detected \citep{Papitto_2013Natur}. 
This source was previously identified as the rotation-powered radio pulsar PSR~J1824$-$2452I in the M28 globular cluster \citep{Manchester_2005AJ_ATNF}. While IGR~J18245$-$2452 was observed during an X-ray outburst, where the high X-ray luminosity ($L_\mathrm{X} \sim 10^{36}-10^{37} \, \mathrm{erg \, s^{-1}}$) enabled the detection of coherent millisecond pulsations, such pulsed signals have never been blindly discovered in the sub-luminous disk state so far. The X-ray luminosity in this state ($L_\mathrm{X} \sim 10^{33}-10^{34} \, \mathrm{erg \, s^{-1}}$), much weaker than that of accreting MSPs in outburst, hinders the discovery of signals at the NS spin period. The detection of X-ray pulsations in the sub-luminous disk state for J1023 \citep{Archibald_2015} and XSS~J12270$-$4859 \citep{Papitto_2015MNRAS} effectively relied on prior knowledge of the pulsar ephemeris obtained during their rotation-powered states \citep{Archibald_2009Sci, Roy2015}.

In light of the prevailing scarcity and sparsity of multi-wavelength observations of J1544, we coordinated the most extensive high-time resolution campaign ever conducted on this source, covering from the radio band to the X-rays.
The goal was to investigate the properties of J1544 in detail across the electromagnetic spectrum, connecting spectral variability and mode-switching phenomena across multiple bands to provide a more comprehensive understanding of the source behavior.

The paper is structured as follows: in Sect.~\ref{sec:obs_J1544}, we present the observations and data analysis. The results are then reported in Sect.~\ref{sec:results}. In Sections \ref{sec:discussion} and \ref{sec:conclusions}, we discuss our findings and draw conclusions, respectively.

\section{Observations and data analysis} \label{sec:obs_J1544}
Table~\ref{tab:log} provides a list of the observations performed during the four-day-long multi-wavelength campaign. 
The first day (Day 1) involved \textit{XMM-Newton}, the fast optical photometer SiFAP2 mounted on the 3.6-m Telescopio Nazionale Galileo (TNG), the \textit{Hubble} Space Telescope (\textit{HST}), the Rapid Eye Mount (REM) telescope, and the Karl G. Jansky Very Large Array (VLA); the second and third days (Day 2-3) involved the Nuclear Spectroscopic Telescope Array (\textit{NuSTAR}) and REM; the fourth day (Day 4) involved the Neutron Star Interior Composition Explorer Mission (NICER), the high-speed, quintuple-beam camera HiPERCAM mounted on the 10.4-m Gran Telescopio Canarias (GTC), and REM. To further characterize the source across different wavelengths, we also included observations acquired with the Near Infrared Camera Spectrometer (NICS) at the TNG in 2016, as well as additional radio data collected in November and December 2024 using the Australia Telescope Compact Array (ATCA). In the following, we detail these observations and outline the procedures used for data processing and analysis. Archival near-infrared (NIR) observations are described in Sect.~\ref{sec:TNG_NICS_data}, while additional radio data acquired in November and December 2024 are detailed in Sect.~\ref{sec:ATCA}.

\begin{table*}
\small
\caption{
\label{tab:log}
Observation log of J1544 in February 2024.} 
\centering
\begin{tabular}{lcccc}
\hline\hline
Telescope/Instrument    & Obs. ID/Project       & Start -- End time                       & Exposure  & Band (setup) \\
                                      &                       & MMM DD hh:mm:ss (UTC)                   & (ks)      &   \\
\hline
\multicolumn{5}{c}{Day\,1}\\
\hline
\textit{XMM-Newton}/EPIC-pn            & 0923466201       & Feb 8 03:41:01  -- Feb 8 20:59:01   &   60.8   & 0.3--10\,keV (fast timing)   \\
\textit{XMM-Newton}/EPIC-MOS1          & 0923466201       & Feb 8 03:59:59  -- Feb 8 20:58:46   &   59.2   & 0.3--10\,keV (small window)  \\
\textit{XMM-Newton}/EPIC-MOS2          & 0923466201       & Feb 8 03:59:27  -- Feb 8 20:58:46   &   59.2   & 0.3--10\,keV (small window)  \\
\textit{XMM-Newton}/OM                 & 0923466201       & Feb 8 00:47:48  -- Feb 8 20:09:20   &   69.6   & $V$ (fast window)  \\
TNG/SiFAP2              & A48DDT3          & Feb 8 04:58:35  -- Feb 8 06:40:35   & 4.32      & fast timing (white light)\\
\textit{HST} STIS/NUV-MAMA       & 17587            & Feb 8 05:52:48  -- Feb 8 06:28:09   & 2.12       & G230L (fast timing spectroscopy)   \\
REM                     &                  & Feb 8 06:36:08 -- Feb 8 06:42:32    & 0.38 & REMIR camera, $H$-band\\                                                                        
VLA                     & SX222346         & Feb 8 11:42:00  -- Feb 8 15:41:20   & 10.4       & Band $C$ (C configuration)  \\

\hline
\multicolumn{5}{c}{Day\,2-3}\\
\hline
\textit{NuSTAR}/FPMA/FPMB       &   92346          & Feb 9 04:10:15  -- Feb 10 06:31:06   & 45.5/45.0 & 3--79\,keV   \\
REM                     &                  & Feb 9 06:31:52 -- Feb 9 06:38:16    & 0.38 & REMIR camera, $H$-band\\                                             
REM                     &                  & Feb 10 05:39:43 -- Feb 10 05:46:07    & 0.38 & REMIR camera, $H$-band\\                                                                        

\hline
\multicolumn{5}{c}{Day\,4}\\
\hline
NICER/XTI               &   6030120103     & Feb 11 06:11:11 -- Feb 11 06:15:11 & 0.24      &  0.2--12\,keV \\
GTC/HiPERCAM            &   GTC129-23B     & Feb 11 04:48:36 -- Feb 11 06:49:55 & 7.3       & $u_s$,$g_s$,$r_s$,$i_s$,$z_s$\\
REM                     &                  & Feb 11 05:39:47 -- Feb 11 05:46:11    & 0.38 & REMIR camera, $H$-band\\

\hline
\end{tabular}
\end{table*}

\subsection{XMM-Newton (Day 1)} \label{sec:XMM_data_analysis}
We analyzed \textit{XMM-Newton} \citep{Jansen_2001AA, Schartel_2022} Target of Opportunity (ToO) observations of J1544 performed on 2024 February 8 (AO-22 Proposal 092346, PI: Miraval Zanon), using the European Photon Imaging Cameras (EPIC) and the Optical/UV Monitor telescope (OM; \citealt{Mason_2001AA}). 
The EPIC-pn \citep{Struder_2001AA} operated with a time resolution of 29.5 $\mathrm{\mu s}$ (fast timing mode), the two EPIC-MOS \citep{Turner_2001AA} with a time resolution of 0.3 s (small window mode), and the OM with a time resolution of 0.5 s (fast window mode) with the V filter (effective wavelength 5407 \AA; full-width at half-maximum 684 \AA). We processed and analyzed the Observation Data Files using the Science Analysis Software (SAS; v.21.0.0). 
We identified high background flaring activity in the 10–12 keV light curves of the whole field of view over a time span of $\sim$4 ks during the initial part of the observation.
However, we checked that these flares did not impact the identification of X-ray modes, hence we decided to keep all the data for the following analyses.
We extracted the spectra as described below, either by discarding the high-background contaminated data or by including it, and verified that the results were entirely consistent with each other.

\begin{figure}
   \centering
   \includegraphics[width=0.5\textwidth]{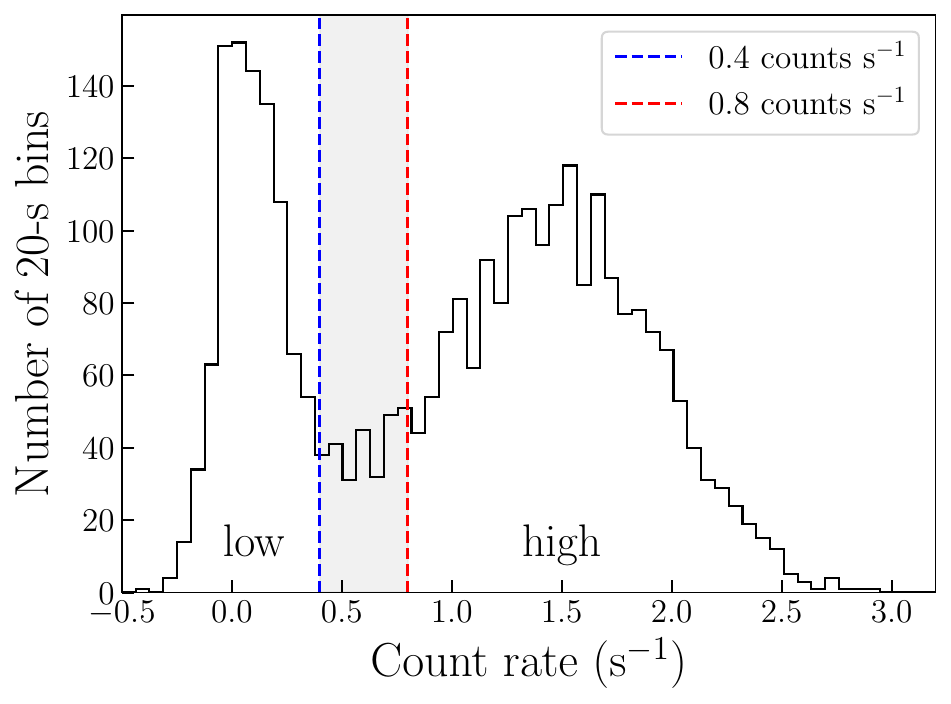}
   \caption{Distribution of count rates obtained from the background-subtracted \textit{XMM-Newton}/EPIC light curve acquired on 2024 February 8, binned with a time resolution of 20 s. We defined the high modes as the time intervals when the count rate exceeded 0.8 counts s$^{-1}$ (red dashed line), and the low modes as those when the count rate dropped below 0.4 counts s$^{-1}$ (blue dashed line). A ``transition'' region is also present, following the procedure adopted by \citet{Bogdanov_2015ApJ_J1023} for J1023. No flaring mode was observed.} \label{fig:bimodality}
\end{figure}

For the EPIC-pn we defined the source region as a 11-pixel-wide strip centered on the brightest pixel column (RAWX=32-42), and the background region as a strip with the same width far from the source position (RAWX=3-13). For each MOS, we extracted source photons within a circular region centered on the source position with a 40\arcsec\, radius, and background photons from a 80\arcsec\, wide, source-free circular region on one of the outer CCDs.
Using the task \texttt{epiclccorr}, we extracted the background-subtracted light curves from the three EPIC instruments over the time interval of simultaneous coverage, binning it with a time resolution of 20 s. We selected the high modes as the time intervals when the count rate exceeded 0.8 counts s$^{-1}$, and the low modes as those when the count rate dropped below 0.4 counts s$^{-1}$ (see Fig.~\ref{fig:bimodality}). No flaring mode was observed. The good time intervals (GTIs) corresponding to the high and low modes were then used to extract background-subtracted spectra, response matrices, and ancillary files separately for each mode.
The average EPIC-pn spectrum was extracted with a minimum of 100 counts in each channel within the 0.7-7.0 keV energy range since EPIC-pn data in timing mode are not well calibrated below 0.7 keV\footnote{\url{https://xmmweb.esac.esa.int/docs/documents/CAL-TN-0018.pdf}.} along with the background domination at energies above 7 keV. For the average EPIC-MOS1 and EPIC-MOS2 spectra extracted in the 0.3-10 keV band, a minimum of 50 counts per channel bin was selected. During high-mode intervals, EPIC spectra were extracted with a minimum of 50 counts per channel bin. Conversely, during low-mode intervals, where the EPIC-pn spectrum is background-dominated, only EPIC-MOS1 and EPIC-MOS2 spectra with a minimum of 20 counts per channel were considered.

We extracted the optical background-subtracted light curve using the \texttt{omfchain} pipeline with default parameters. The light curve was binned at a time resolution of 1200 s to search for any orbital modulation as previously done by \citet{Bogdanov_2015ApJ}.
We converted the observed count rates, $\mathrm{R_{OM}}$, to magnitudes in the Vega system using the relation\footnote{\url{https://www.cosmos.esa.int/web/xmm-newton/sas-watchout-uvflux}.} mag=$17.963-2.5 \, \log_{10}(\mathrm{R_{OM}})$.
Following the procedure described by \citet{Baglio_CotiZelati_2023A&A} to align the X-ray and optical light curve with those obtained from other observations at different wavelengths, we converted the photon arrival times from the Terrestrial Time (TT) to the Coordinated Universal Time (UTC) without applying a barycentric correction\footnote{Given a ground-based telescope positioned on the opposite side of the Earth from a satellite, and accounting for both Earth's diameter and various telescope altitudes, the optical path difference is $\sim$44 ms for NICER, \textit{NuSTAR}, and \textit{HST}, and in the range of $\sim$61-419 ms for \textit{XMM-Newton}. Since the light curves are binned at 10\,s for NICER, 20\,s for both \textit{XMM-Newton} and \textit{HST}, and 100\,s for \textit{NuSTAR}, this path difference does not introduce any relevant time shifts for the analyses in this work.}.

\subsection{TNG/SiFAP2 (Day 1)}
The fast optical photometer SiFAP2 \citep{Ghedina_2018SPIE}, mounted on the 3.58-meter INAF Telescopio Nazionale Galileo (TNG; \citealt{Barbieri_1994SPIE}), observed J1544 on 2024 February 8 (PI: Illiano), simultaneously with \textit{XMM-Newton} and \textit{HST} observations.
SiFAP2 observations were performed in white light spanning the 320–900 nm band and peaking between 400 and 600 nm (i.e., roughly corresponding to the B and V Johnson ﬁlters; see Supplementary Figure 1 in \citealt{Ambrosino_2017}).
Due to adverse weather conditions and gusty winds, we only retained data between 60348.21 MJD and 60348.26 MJD, corresponding to a total duration of 4320 s.
To take into account spurious sky effects and monitor the sky condition, we simultaneously observed a V=12.3 mag reference star TYC 5605-459-1 \citep{Hog_2000A&A}.
We estimated the sky background by offsetting the telescope in a region located 10\arcsec\, away from the target (i.e., outside the $7$\arcsec $\times 7$\arcsec\, region observed when pointing at the source position) towards the east direction for 31\,s at the start of the observation. The background rates are 17249.8 counts\,s$^{-1}$ for the source channel and 24568.3 counts\,s$^{-1}$ for the reference channel, the difference being due to the presence of clouds.
During the retained time interval, the mean source count rate was 13837.5 counts\,s$^{-1}$, with an extrapolated background count rate of 12848.3 counts\,s$^{-1}$. For the reference channel, the mean count rate was 581086.8 counts\,s$^{-1}$, with an extrapolated background of $\sim$$2.3 \times 10^{4}$ counts\,s$^{-1}$. 
To extract a background-subtracted light curve, the background counts are subtracted from the source counts. This process often resulted in counts consistent with zero within the errors, indicating that the source was not detected. 

\subsection{HST (Day 1)} \label{sec:HST_analysis}
The Space Telescope Imaging Spectrograph (STIS; \citealt{Woodgate_1998PASP}) aboard the \textit{Hubble} Space Telescope (\textit{HST}) observed J1544 on 2024 February 8 for $\sim$35 minutes (Program 17587, PI: Illiano). These observations were performed using the near-UV Multi-Anode Micro-channel Array detector (NUV-MAMA) in TIME-TAG mode (time resolution of 125 $\mu s$), and collected with the G230L grating equipped with a 52\arcsec $\times$ 0.2\arcsec slit with a spectral resolution of $\sim$500 over the nominal range of $1570-3180$ \r{A}.
Due to high background noise in the spectrum outside the $2000-3000$ \r{A} range, we restricted our spectral and timing analysis to this interval. However, we verified that the light curve does not change significantly when comparing the nominal and chosen ranges.
We calibrate the UV data using the CALSTIS v3.4.2 pipeline, which includes dark subtraction and flat fielding. To extract the background-subtracted light curve, we employed the \texttt{stistools} Python package’s \texttt{inttag} function\footnote{\url{https://stistools.readthedocs.io/en/latest/inttag.html}.} to obtain 20-s-long time series before applying CALSTIS.

\subsection{REM (Day 1-2-3-4)} \label{sec:REM_analysis}
The Robotic Eye Mount (REM) Telescope located in La Silla, Chile, observed J1544 over four consecutive nights: 2024 February 8, 9, 10, and 11 (MJD 60348 -- 60351; PI: Baglio).
For each night, three sets of observations were acquired using the near-infrared (NIR) camera with the $H$-band filter (central wavelength: 1.662 $\mathrm{\mu m}$). Each set comprised five dithered 30-second integration exposures, combined to enhance background subtraction.
The target was not bright enough to be detected in individual exposure images. Additionally, seeing conditions were suboptimal, with the PSF varying between $2''$ and $4''$ across different epochs.
To attempt a detection, we averaged all images for each epoch. After averaging, the target was consistently detected, except for the final epoch on February 11, when the seeing conditions were at their worst ($\sim 4''$). 
We conducted aperture photometry using the {\tt daophot} tool \citep{Stetson_1987PASP}, and an aperture approximately twice the average full-width at half-maximum (FWHM) of the flux distribution of field stars. Magnitudes were then calibrated against a group of four bright stars in the field, with magnitudes reported in the Two-Micron-All-Sky-Survey (2MASS; \citealt{Skrutskie_2006AJ})\footnote{\url{https://irsa.ipac.caltech.edu/Missions/2mass.html}.} catalogue.

\subsection{VLA (Day 1)} \label{sec:VLA_analysis}
The Karl G. Jansky Very Large Array (VLA) observed J1544 (project code: SX222346, PI: Miraval Zanon) on 2024 February 8, for a total of 4\,hr, at the central frequency of 6\,GHz (C-band), with a bandwidth of 4\,GHz. The target and the phase calibrator J1543$-$0757 were observed in ten-minute cycles, with eight minutes on the former and two minutes on the latter. The distance between the target and the phase calibrator is about 3.5$^\circ$. The observation included a scan on the flux and bandpass calibrator (J1331$+$3030). The total on-source time is $\sim$2.89 hr (specifically 10395 s), corresponding to $\sim$72\% of the total observation time, while the remaining 28\% was allocated to calibrators and overheads.
Data were calibrated using the custom \texttt{casa} pipeline \citep[Version 6.5.4,][]{casa2} and visually inspected for possible radio frequency interference (RFI). The final images were produced with the \texttt{tclean} task in \texttt{casa} (Version 6.5.4), using Briggs weighting with robust parameter 0.5.

\subsection{NuSTAR (Day 2-3)} \label{sec:NuSTAR_data_analysis}
\textit{NuSTAR} \citep{Harrison_2013ApJ} observed J1544 on 2024 February 9, for a total elapsed time of $\sim$94.8 ks and a net exposure of $\sim$45 ks (as part of the joint \textit{NuSTAR}/\textit{XMM-Newton} cycle-22 program 092346, PI: Miraval Zanon)\footnote{These observations were scheduled to take place simultaneously with GTC/HiPERCAM optical observations on three slots on February 9, 10, and 11 from 04:30 to 06:30 UTC. However, the \textit{NuSTAR} observations provided coverage only during the first two slots due to a high-urgency ToO observation that took precedence during the third slot. On the other hand, no optical observation was performed during the first two slots owing to unfavorable weather conditions. Therefore, unfortunately, the \textit{NuSTAR} observations did not take place simultaneously with any other observation.}.
Data were processed and analyzed using NuSTARDAS v2.1.2 along with the calibration database (CALDB) 20240506. We accounted for South Atlantic Anomaly passages using \texttt{nupipeline} task with \texttt{SAAMODE = optimize} and \texttt{TENTACLE = yes}.
Source and background events were extracted from circular regions of 60\arcsec\, radius centered at the source position and in a source-free area of the CCD, respectively. J1544 was detected up to energies of $\sim$30 keV in the two focal plane modules
(FPMA and FPMB). Light curves were extracted separately for the two modules, combined to increase the Signal-to-Noise ratio (SNR), and binned at a time resolution of 100 s.
Spectra for both detectors and the corresponding response files were extracted using the \texttt{nuproducts} command. We grouped the average spectra with a minimum of 50 counts per channel. Spectra were also extracted separately during high-mode intervals, ensuring a minimum of 20 counts per channel. For the low-mode intervals, the counting statistics was insufficient to extract meaningful spectra.

\subsection{NICER (Day 4)} \label{sec:NICER_data_analysis}
NICER \citep{NICER_Gendreau_2012} observed J1544 on 2024 February 11 (ObsID 6030120103, PI: Illiano).
We analyzed the data using HEASoft version 6.33.2 and NICERDAS version 12, 
with CALDB version 20240206. After retrieving the latest geomagnetic data with the \texttt{nigeodown} task, we processed the observational data using \texttt{nicerl2} task with default screening parameters. 
The obtained cleaned event files were used to generate the light curve in the 0.5--10\,keV energy range with a time bin of 10\,s using the \texttt{nicerl3-lc} pipeline with \texttt{SCORPEON} background model\footnote{See \url{https://heasarc.gsfc.nasa.gov/docs/nicer/analysis_threads/scorpeon-overview/}.}, version 23.
Due to visibility issues, only 240 s of NICER observations were conducted simultaneously with GTC/HiPERCAM, and no clear low-mode intervals were detected during this limited timeframe. Consequently, we opted not to include the NICER data in the subsequent analysis.

\subsection{GTC/HiPERCAM (Day 4)}
The high-speed, quintuple-beam camera HiPERCAM \citep{Dhillon_2021MNRAS_HiPERCAM} mounted on the 10.4-m Gran Telescopio Canarias (GTC) on the island of La Palma (Canary Islands) observed J1544 for $\simeq$2\,hr on 2024 February 11 (PI: Coti Zelati). HiPERCAM was operated in windowed mode to achieve single frame exposure times of $\simeq$0.19\,s for the Super SDSS $g_S$, $r_S$, $i_S$, and $z_S$ filters and three times longer in the $u_S$ filter (to compensate for the lower filter throughput). The average seeing during the observations was $\simeq 1.7''$. 

We processed and analyzed the data using the HiPERCAM reduction software v. 1.5.2\footnote{\url{https://cygnus.astro.warwick.ac.uk/phsaap/hipercam/docs/html/index.html}.}.
We processed each science frame by subtracting the bias and applying a flat-field correction using the median of twilight-sky frames. Additionally, we corrected for potential fringing effects in the $z_S$ filter by using publicly available, pre-prepared fringe maps. We used variable aperture photometry to extract counts from the target and a bright comparison star, that is, for each frame we adjusted the aperture radius based on the FWHM of the fitted profile of the comparison star.
The comparison star was Gaia DR3 6268482198463636224 (also referred to as USNO-B1.0 0785-0287404).
We estimated the sky level around the target and the comparison star by averaging the counts within an annular region surrounding each of them. Next, we performed differential photometry by dividing the background-subtracted counts of J1544 by those of the comparison star. Finally, the time series were rebinned to a time bin of 20\,s to increase the SNR while still allowing us to study the variability of the optical emission over time scales compatible with those expected for the X-ray mode switching. 

\subsection{TNG/NICS (2016)} \label{sec:TNG_NICS_data}
To further characterize J1544 across different wavelengths, we included observations acquired with the Near Infrared Camera Spectrometer (NICS; \citealt{Baffa_2001A&A}) at the TNG. NICS observations were conducted during two nights on 2016 March 29 and 30 (Prg. ID: A33DDT4A; PI: de Martino) for a total exposure of $\sim$4.8\,hr and $\sim$4.9\,hr respectively, under good seeing conditions. The J, H, and K filters were used cyclically, with 20-s exposures at 10 mosaic positions.
The image reduction was performed using the {\em jitter} task of the ESO-eclipse software package\footnote{\url{https://www.eso.org/sci/software/eclipse/}.}. 
Aperture photometry was carried out using the {\em IRAF} {\sc DAOPHOT} package \citep{Stetson_1987PASP}. The magnitudes were calibrated with the two nearby stable reference stars listed in the 2MASS catalog, 2MASS15444507-11284959 and 2MASS15443354-112930.

\subsection{ATCA (Nov./Dec. 2024)} \label{sec:ATCA}
We obtained additional radio observations of J1544 with the Australia Telescope Compact Array (ATCA) under program CX588 (PI: Carotenuto). The first observation was conducted on 2024 November 25 from 01:40 UT to 07:00 UT, with the telescope in the 750D configuration. A second observation was carried out from 2024 December 3 at 20:40 UT to 2024 December 4 at 02:15 UT, with the telescope in the H214 configuration.
For both epochs, data were recorded simultaneously at central frequencies of 5.5\,GHz and 9.0\,GHz, with 2\,GHz of bandwidth at each frequency. We used PKS~B1934$-$638 for the bandpass and flux density calibration, and PKS B1540$-$077 for the complex gain calibration. Data were flagged, calibrated, and imaged using standard procedures within {\tt CASA}. 
When imaging, we included antenna CA06 for both epochs, and we used a natural weighting scheme to minimize the root mean square (RMS) noise of the image.

\begin{figure*}
    \centering
    \includegraphics[width=1\textwidth]{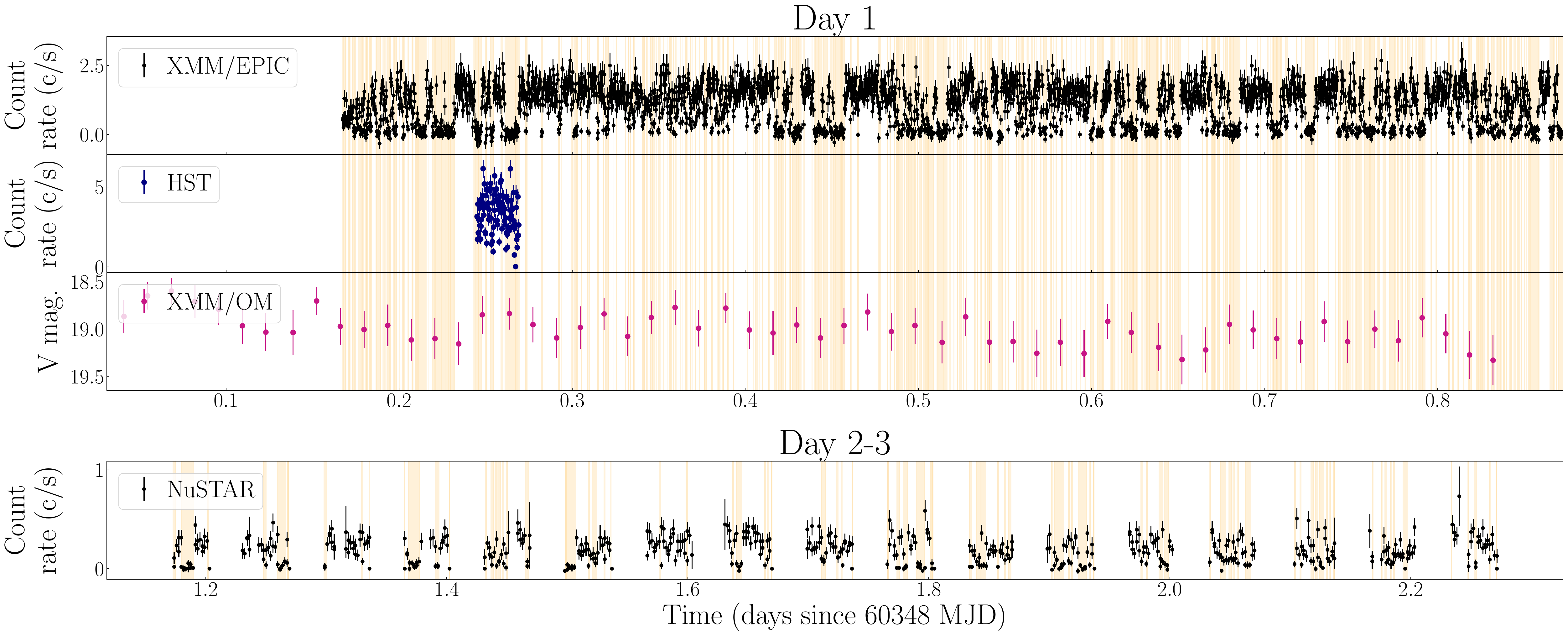}
    \caption{Temporal evolution of the X-ray, UV, and optical emissions of J1544 during the first three days of observations. 
    For Day 1, the light curves are shown in decreasing order of energy band, from top to bottom, with the \textit{XMM-Newton}/EPIC (20 s time bin), \textit{HST}/STIS (20 s), and \textit{XMM-Newton}/OM (1200 s). For Days 2–3, the 3-30~keV \textit{NuSTAR} light curve (100 s time bin) is shown. The yellow-shaded areas denote the time intervals of the low X-ray mode identified by \textit{XMM-Newton}/EPIC on Day 1 and by \textit{NuSTAR} on Day 2-3. The error bars represent 1$\sigma$ uncertainties.}\label{Fig:light_curves}
\end{figure*}

\begin{figure*}
    \centering
    \includegraphics[width=1\textwidth]{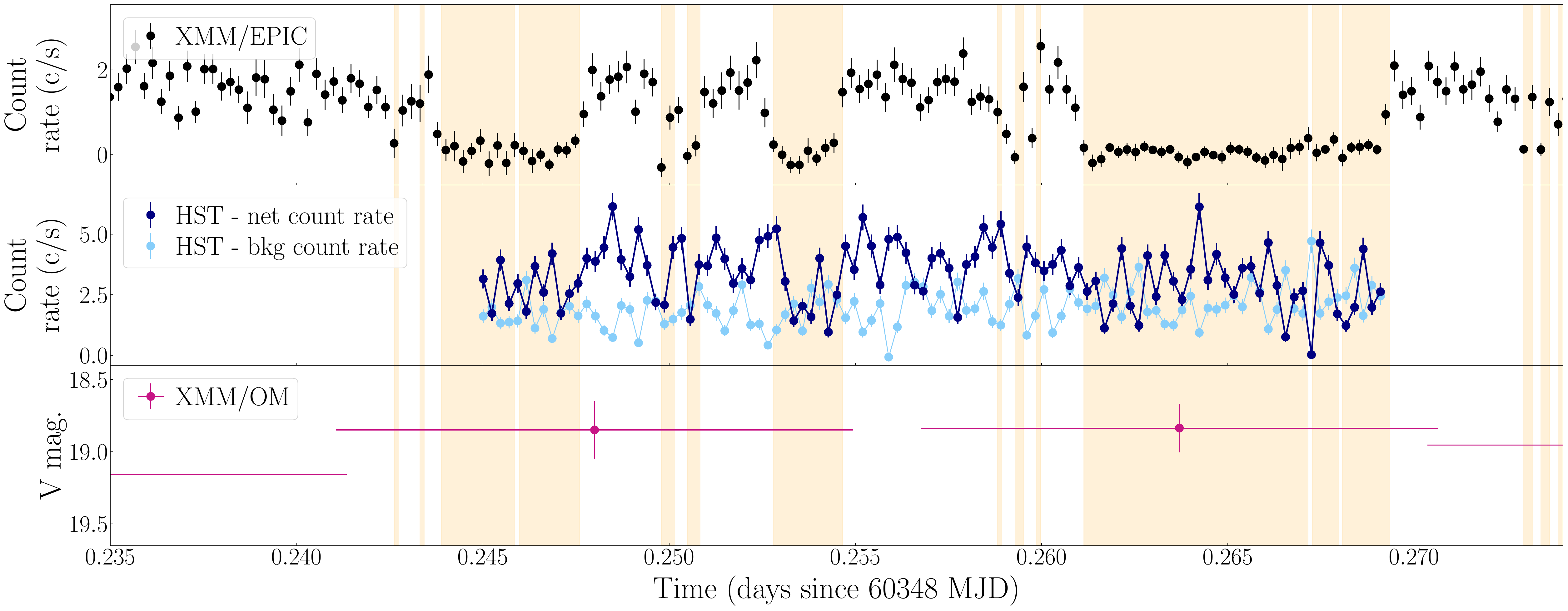}
    \caption{Zoom-in of the time series collected on Day 1, capturing the period of maximum overlap among different telescopes.}\label{Fig:light_curves_zoom}
\end{figure*}

\section{Results} \label{sec:results}

\subsection{X-ray emission} \label{sec:X-ray_emission}
The \textit{XMM-Newton}/EPIC background-subtracted light curve extracted during Day 1 revealed hundreds of low-mode intervals (see Fig.~\ref{Fig:light_curves}).
By applying our threshold to discern different intensity levels in the EPIC net count rates (see Sect.~\ref{sec:XMM_data_analysis}), we determined that J1544 spent $\simeq$58\% and $\simeq$33\% of the observation time in the high and low modes, respectively (similar proportions were reported by \citealt{Gusinskaia_2025MNRAS}), and the remaining time switching between these modes. 
The shortest observed duration of a low-mode episode spanned $\sim$20\,s (corresponding to a single time bin), whereas the longest recorded duration extended to $\sim$520\,s (occurring between $\sim$60348.26~MJD and $\sim$60348.27~MJD; see Fig.~\ref{Fig:light_curves_zoom}).

The \textit{NuSTAR} light curve displays a lower SNR compared to the \textit{XMM-Newton}/EPIC light curve (see Fig.~\ref{Fig:light_curves}). However, we identified high- and low-mode intervals and used the GTIs to extract background-subtracted spectra, response matrices, and ancillary files separately for each mode.

We performed the X-ray spectral analysis on \textit{XMM-Newton} and \textit{NuSTAR} data (see Sects.~\ref{sec:XMM_data_analysis} and \ref{sec:NuSTAR_data_analysis}) using the X-ray spectral ﬁtting package XSPEC \citep{Arnaud_1996_XSPEC} version 12.14.0. We adopted the interstellar medium abundance and the cross-section tables from \citet{Wilms_2000ApJ} and \citet{Verner_1996ApJ}, respectively. Uncertainties on spectral parameters are given at 1$\sigma$ confidence intervals unless otherwise stated.
The average broadband spectrum is well described by an absorbed power-law model (\texttt{constant * TBabs * powerlaw}). We included a renormalization factor in our spectral modeling to address cross-calibration uncertainties between the two X-ray telescopes. We verified that these factors were consistent within 10\% for both the average broadband spectrum and the spectra extracted separately in the high- and low-mode intervals. This condition was satisfied in all cases except for \textit{NuSTAR} in the average spectrum, where the constant deviated by 11\%.  The discrepancy can be attributed to the \textit{NuSTAR} observations being conducted one day after the \textit{XMM-Newton} observations, resulting in differences in the number and duration of the high and low modes and thus the average intensity level. Supporting this interpretation, the renormalization factors for the spectra extracted separately in the high- and low-mode intervals remained consistent within 10\%.
The best-fitting parameters are an absorption column density of $\mathrm{N_H} = (1.52\pm0.06) \times 10^{21} \, \mathrm{cm^{-2}}$, a photon index of $\Gamma = 1.63 \pm 0.01$, and an unabsorbed 0.3-10 keV flux of $F_{\mathrm{unabs}} = (3.54 \pm 0.02) \times 10^{-12} \, \mathrm{erg \, cm^{-2} \, s^{-1}}$ ($\chi^2$ = 445.7 for 414 degrees of freedom, d.o.f.). 
The estimated value of $\mathrm{N_H}$ is broadly consistent with the average absorption column density in the direction of the source of $\sim$$1.2 \times 10^{21} \, \mathrm{cm^{-2}}$ \citep{HI4PICollaboration_2016A&A}, and both $\mathrm{N_H}$ and $\Gamma$ are consistent within 3$\sigma$ with values previously reported for this source from X-ray observations by \citet{Bogdanov_2015ApJ}, \citet{Bogdanov_2016ApJ}, and \citet{Gusinskaia_2025MNRAS}.
Given that the reflection spectrum, and in particular the iron K$\alpha$ complex, is often observed in LMXBs hosting a NS \citep[see, e.g.,][and references therein]{DiSalvo_2023hxga.book}, we searched for similar features in J1544 by adding Gaussian terms to the power-law model. This approach is analogous to that used for the search in other tMSPs or candidate systems in the sub-luminous disk state. The absorption column density and the photon index of the power law were fixed to their best-fit values from the previous analysis.
The line energies were set sequentially at 6.4 keV (for neutral or lightly ionized Fe I K$\alpha _1$ and K$\alpha _2$ transitions), then 6.7 keV (He-like Fe XXV), and 6.97 keV (H-like Fe XXVI). To simulate lines narrower than the instruments' intrinsic energy resolution, we set the Gaussian line width to zero. No significant emission feature was detected. As a result, we determined 3$\sigma$ upper limits for the equivalent widths of any Gaussian features at 6.4, 6.7, and 6.97 keV to be 20, 23, and 32 eV, respectively. The iron line has not been detected in this system or any other confirmed or candidate tMSPs in the sub-luminous disk state, including J1023 \citep[see, e.g.,][]{CotiZelati2014, CotiZelati_2019A&A} and XSS~J12270$-$4859 \citep{deMartino_2010A&A}, providing no evidence for fluorescence or disk reflection in these systems.

When fitting the spectra extracted separately in the high- and low-mode intervals, we fixed the $\mathrm{N_H}$ to the value obtained from the average spectral fitting. 
The high-mode spectrum includes all \textit{XMM-Newton} and \textit{NuSTAR} data, whereas the low-mode spectrum comprises only the two MOS data due to background dominance in the EPIC-pn and low statistics in \textit{NuSTAR} data.
For the high mode, we obtained $\Gamma_{\rm H} = 1.623\pm0.008$, and unabsorbed 0.3-10\,keV flux of $F_{\mathrm{unabs,H}} = (5.58 \pm 0.04) \times 10^{-12} \, \mathrm{erg \, cm^{-2} \, s^{-1}}$
($\chi^2$/d.o.f =521.5/538), while for the low mode $\Gamma_{\rm L} = 1.65 \pm 0.07$ and an unabsorbed 0.3-10\,keV flux $F_{\mathrm{unabs, L}} = (0.38 \pm 0.02) \times 10^{-12} \, \mathrm{erg \, cm^{-2} \, s^{-1}}$ ($\chi^2$/d.o.f =37.9/45). The values for the photon indices in the different modes are compatible with each other and with the average value within 1$\sigma$. To improve the statistics and search for a potential steepening of the photon index in the low mode \citep{Campana_2019A&A}, we added all \textit{XMM-Newton} archival observations performed on this source at the moment of writing. The results remained consistent with those presented above, with a slight improvement in parameter precision, as detailed in Appendix~\ref{Sec:appendix_spectral_analysis}.

\subsection{UV emission}
We conducted \textit{HST} observations on Day 1 concurrently with \textit{XMM-Newton} to investigate the presence of a bimodal pattern in the UV data. Notably, similar switches between high and low modes have been recorded in \textit{HST} data of J1023 \citep{Miraval_Zanon_2022, Jaodand2021, Baglio_CotiZelati_2023A&A}. The UV light curve of J1544, shown in Fig.~\ref{Fig:light_curves_zoom} (dark blue points), exhibits substantial variability, with some instances appearing to align with the varying intensity levels observed in X-rays, particularly between $\sim$60348.25 and $\sim$60348.26 MJD.
However, the background level (light blue points in Fig.~\ref{Fig:light_curves_zoom}) varies in amplitude and timescale similarly to the total count rate, particularly in the final part of the \textit{HST} observation. As a result, the expected high/low mode switches, likely weaker than this variability, may be obscured by background contamination.
Future \textit{HST} observations may help determine whether this background variability is persistent and, despite the faint nature of J1544, potentially enable us to identify bimodality within the count rate distribution.

\subsection{Optical variability and its connection with mode switching} \label{sec:optical_results}
Attempts to bin the \textit{XMM-Newton}/OM light curve acquired on Day 1 using time bins of a few hundred seconds were unsuccessful in detecting mode switching.
As noted by \citet{Bogdanov_2015ApJ}, the faint nature of J1544 makes it challenging to identify variability on timescales of a few tens of seconds with the OM. 
Even for J1023, variability on such short timescales was not observed, likely masked by the intrinsic statistical uncertainties of the data \citep{Baglio2019}.
Consequently, we binned the OM light curve using a time bin of 1200 s (see Fig.~\ref{Fig:light_curves}) to increase the SNR.
We observed similar fluctuations in the V magnitude as reported by \citet{Bogdanov_2015ApJ} in the U filter, recording a modulation semi-amplitude of $\sim$0.4 mag.
To quantify the object's inherent variability beyond the limitations imposed by observational uncertainties, we computed the excess variance \citep{Nandra_1997, Edelson_2002, Vaughan_2003, Yuk_2022ApJ}, which resulted to be of $(-0.02 \pm 0.03) \, \mathrm{mag^2}$. This value is consistent with zero within the uncertainty and suggests that the observed variability is not intrinsic to the source. 
The U filter data acquired in 2014 showed a sinusoidal pattern with a periodicity of $\sim$5.2 hr \citep{Bogdanov_2015ApJ}, differing from the orbital period of $\sim$5.8 hr estimated through spectroscopic observations by \citet{Britt_2017ApJ} using radial velocity curves. To further investigate this discrepancy and compare with data in the V filter, we computed the excess variance also on the U filter data, obtaining $(0.01 \pm 0.05) \, \mathrm{mag^2}$. This value is also consistent with zero within the uncertainty, indicating that we cannot draw definitive conclusions.

\begin{figure*}
    \centering
    \includegraphics[width=0.9\textwidth]{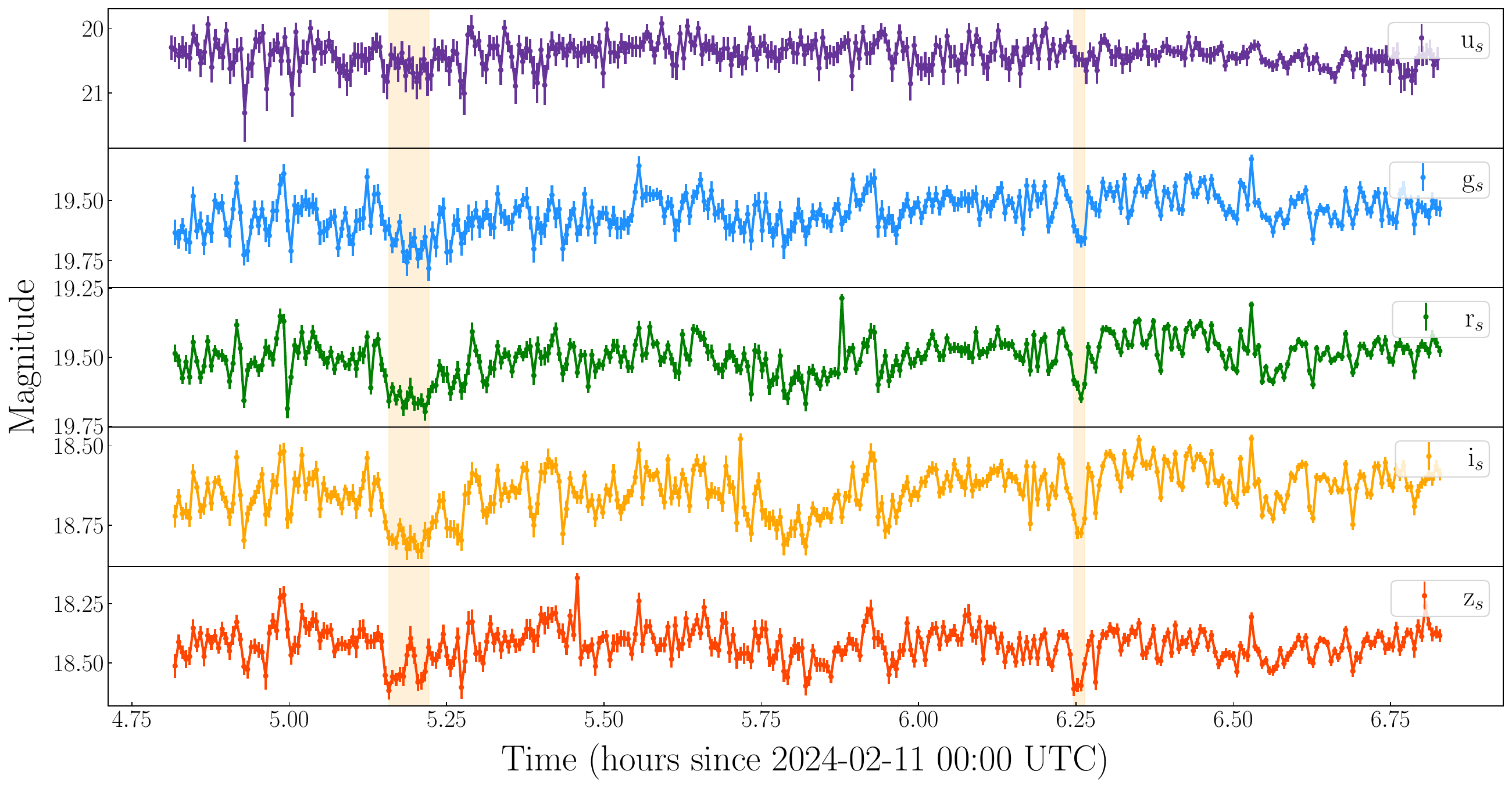}
    \caption{GTC/HiPERCAM light curves acquired on Day 4 in the Super SDSS $u_S$, $g_S$, $r_S$, $i_S$, and $z_S$ filters. The yellow-shaded areas highlight the most prominent potential low modes. The error bars represent 1$\sigma$ uncertainties.}\label{Fig:HiPERCAM_lc}
\end{figure*}

\begin{figure*}
    \centering
    \includegraphics[width=0.9\textwidth]{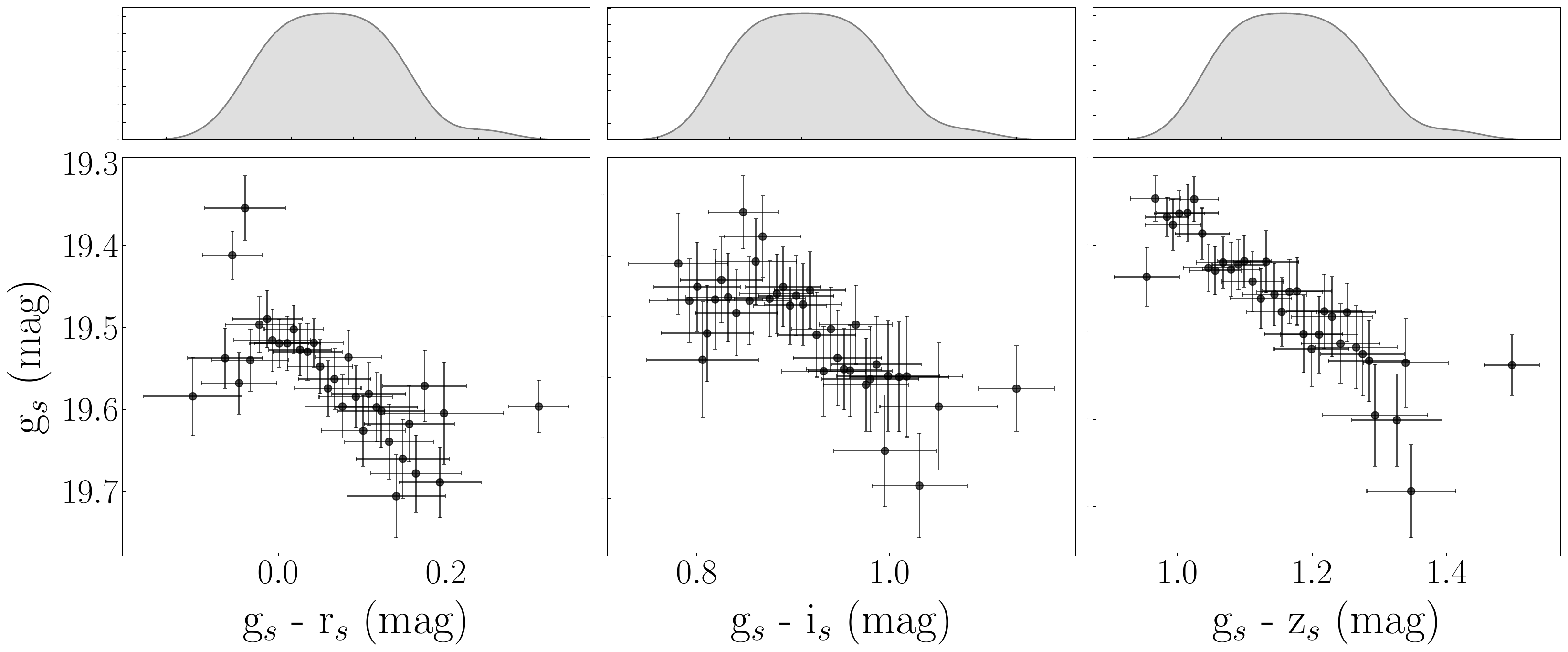}
    \caption{Color-magnitude diagrams for J1544, derived from the most highly sampled $g_S$, $r_S$, $i_S$, and $z_S$ bands of the GTC/HiPERCAM time series (Fig.~\ref{Fig:HiPERCAM_lc}), and rebinned to emphasize correlation trends. The upper panels display the smoothed Kernel Density Estimation curves that illustrate the distribution of the color-magnitude values.}\label{Fig:CMD}
\end{figure*}

GTC/HiPERCAM light curves acquired on Day 4 displayed flickering and dipping activities (Fig.~\ref{Fig:HiPERCAM_lc}) reminiscent of the X-ray bimodal pattern. This optical behavior resembled that observed for the prototype tMSP J1023 \citep{Shahbaz_2015, Shahbaz_2018, Hakala2018} as well as for the candidate CXOU~J110926.4$-$650224 \citep{CotiZelati2024}.
\cite{CotiZelati2024} recently reported that the optical emission from the candidate tMSP CXOU~J110926.4$-$650224 becomes redder during the low mode compared to the high mode, which aligns with the expectation of the proposed model that explains optical and X-ray pulsations from J1023 as a result of synchrotron radiation at the boundary region between the striped pulsar wind and the inner accretion disk \citep{Papitto_2019ApJ, Veledina_2019}. The findings from the extensive multi-wavelength campaign conducted by \citet{Baglio_CotiZelati_2023A&A} on J1023 supported the hypothesis that during switches from high to low modes, the inner and hotter regions of the accretion flow are ejected, resulting in a decline of X-ray, UV and optical fluxes. This leaves a fainter optical emission from the cooler outer regions of the accretion disk. 
Taking into account these recent observations, along with the presence of at least two distinct rectangular flat bottom dips highlighted in Fig.~\ref{Fig:HiPERCAM_lc}, we plotted the magnitudes in the $g_S$ filter against the magnitude colors $g_S-r_S$, $g_S-i_S$, $g_S-z_S$ (see Fig.~\ref{Fig:CMD}).
To quantify the observed correlation, we rebinned our data to reduce the scatter and computed Spearman's rank correlation coefficients, which were found to be $\simeq$0.77, 0.78, and 0.93, respectively, for 32 d.o.f.. The corresponding probabilities of observing such coefficients from uncorrelated data are $\simeq$$8.8 \times 10^{-8}$, $1.3 \times 10^{-8}$, and $1.9 \times 10^{-16}$, respectively. These results indicate that the optical emission becomes significantly redder during the low mode compared to the high mode, particularly when measured using the $g_S$ and $z_S$ filters, which are separated by the largest wavelength intervals.

\subsection{Hint of an optical flare} \label{sec:hint_optical_flare}
\begin{figure*}
    \centering
    \includegraphics[width=0.9\textwidth]{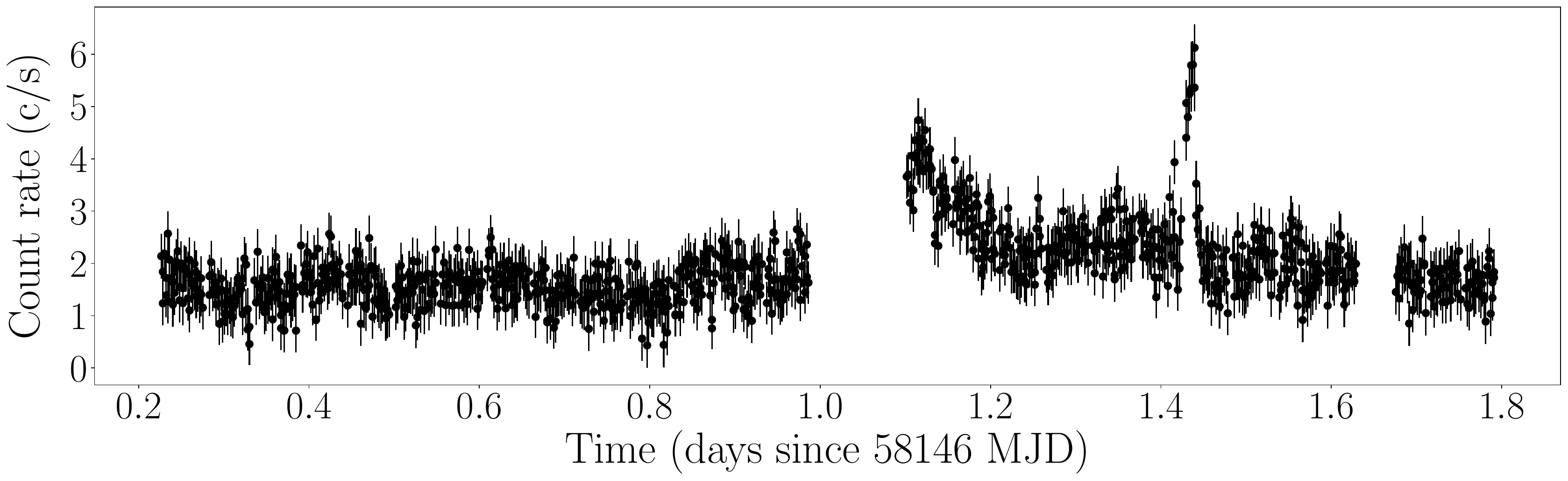}
    \caption{\textit{XMM-Newton}/OM light curve acquired in 2018 using 100-s bins. The error bars represent 1$\sigma$ uncertainties.}\label{fig:optical_flare_2018}
\end{figure*}

While analyzing archival \textit{XMM-Newton} data, we identified peculiar features in the OM light curve acquired on 2018 January 28 (ObsID 0800280101). The OM was operated in fast window mode with a time resolution of 0.5~s in white light. We analyzed the data as detailed in Sect.~\ref{sec:XMM_data_analysis}. As shown in Fig.~\ref{fig:optical_flare_2018}, the OM light curve reveals an increase in flux following an observational gap around 58147.1~MJD and a distinct flare-like feature around 58147.4~MJD. These findings are particularly intriguing, as no flares have previously been reported from J1544, which is peculiar given that confirmed and candidate tMSPs frequently exhibit flaring activity \citep[e.g.,][]{Tendulkar_2014, Bogdanov_2015ApJ_J1023, CotiZelati2024, Li_2020ApJ}. If confirmed, this would represent the first detection of a flare from this source. However, caution is warranted due to the lack of a corresponding feature in the X-ray light curve, as shown in the bottom panel of Fig.~\ref{fig:OM_2018}. Appendix~\ref{sec:appendix_flare} provides a detailed analysis to verify that the observed flare-like event is not spurious or instrumental and to explore its possible origin.

\subsection{Near infrared variability} \label{sec:NIR_variability}
Over the four days of this multi-wavelength campaign, we performed NIR observations using the REM telescope with the $H$-band filter. As outlined in Sect.~\ref{sec:REM_analysis}, due to the source faintness and suboptimal seeing conditions, we were only able to detect J1544 by averaging all exposure images for each observational epoch. On the fourth night, the poor seeing conditions prevented any detection.
The $H$-band magnitude of J1544 was found to vary to some extent between epochs. Specifically, we obtained $H=(16.86\pm 0.26)$~mag, $H=(16.34\pm 0.19)$~mag, $H=(16.73\pm 0.26)$~mag in the first three epochs of observations, respectively, while we determined a 3$\sigma$ upper limit of $>16.88$~mag during the last epoch.
Although the REM observations do not allow for studying short-term variability on timescales of a few tenths of a second, the reported average magnitudes still indicate that the NIR emission is variable.

To further enhance the multi-band characterization of J1544, we analyzed NIR observations acquired in 2016 with TNG/NICS. 
Due to limited X-ray coverage, we cannot definitively confirm that the source was in the intermediate sub-luminous disk state during our NIR observations. However, X-ray flux estimates derived from \textit{Swift}/XRT observations on 2015 October 2 and 2016 May 7 (Obs. IDs: 00084762008 and 00092235001; $\simeq$$(3-4) \times 10^{-12} \, \mathrm{erg \, cm^{-2} \, s^{-1}}$ in the $1-10$~keV energy range) align with the values reported by \citet{Jaodand_2021ApJ} and fall within the expected range for the sub-luminous disk state of tMSPs. Although state transitions can occur on timescales as short as a few weeks \citep[e.g.,][]{Papitto_2013Natur}, the sub-luminous disk state in tMSPs like J1544 has shown remarkable stability over periods exceeding a decade \citep[e.g.,][]{Stappers_2014ApJ, Baglio_CotiZelati_2023A&A}. This sustained stability suggests that J1544 has likely remained in this state since its discovery.

\begin{figure*}
    \centering
    \includegraphics[width=0.9\textwidth]{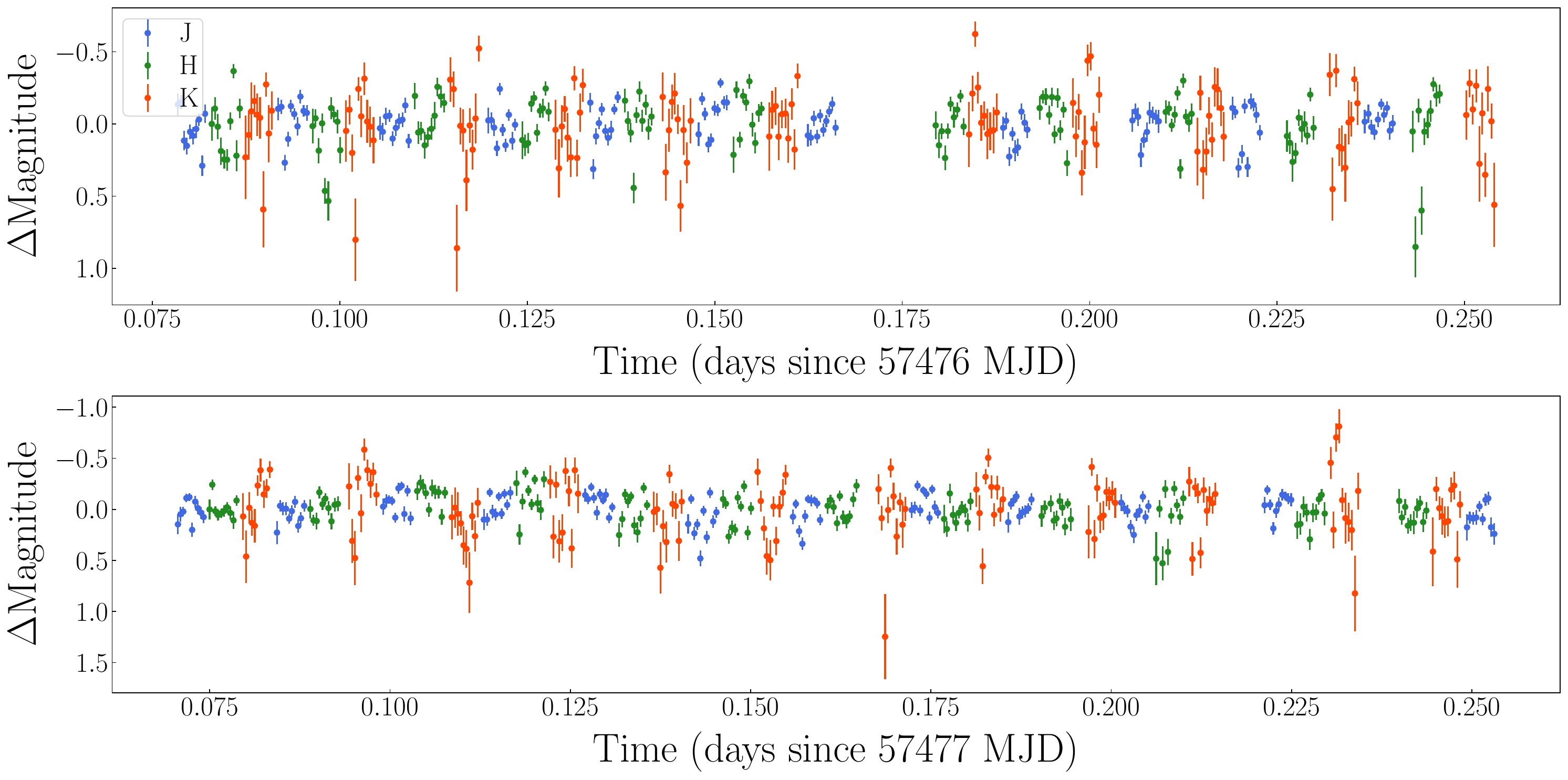}
    \caption{NIR light curves of J1544 observed with TNG/NICS on 2016 March 29 (\emph{top panel}) and 30 (\emph{bottom panel}). We plotted differential magnitudes by subtracting the weighted average value in each filter for each night (see Sect.~\ref{sec:NIR_variability} for details). Photometric data were acquired in the J (blue points), H (green points), and K (red points) filters. The error bars represent 1$\sigma$ uncertainties.}\label{Fig:lc_nics}
\end{figure*}

Figure~\ref{Fig:lc_nics} shows the light curves alternately acquired in the J, H, and K filters. The NIR time series exhibit a variability that may resemble the X-ray bimodal pattern. To highlight this, we plotted the differential magnitudes by subtracting the weighted average value in each filter for each night. The weighted average magnitudes of J1544 during the first night were J$=(17.083 \pm 0.005)$ mag, H$=(16.659 \pm 0.006)$ mag, and K$=(16.32 \pm 0.01)$ mag, and during the second night J$=(17.100 \pm 0.005)$ mag, H$=(16.644 \pm 0.006)$ mag, and K$=(16.29 \pm 0.01)$ mag. The consistency between these average magnitudes in the H band and those observed by REM during our 2024 multi-wavelength campaign further supports that J1544 was in the sub-luminous disk state during the TNG/NICS observations in 2016.
However, the lack of simultaneous X-ray coverage prevented us from confirming the presence of high and low modes in the NIR band. We computed the excess variance, as outlined in Sect.~\ref{sec:optical_results}, to quantify the intrinsic variability beyond measurement uncertainties.  
The estimated values for the J, H, and K bands during the first night are $(0.01 \pm 0.02) \, \mathrm{mag^2}$, $(0.03 \pm 0.03) \, \mathrm{mag^2}$, and $(0.04 \pm 0.03) \, \mathrm{mag^2}$, respectively. During the second night of NIR observations, we obtained $(0.01 \pm 0.02) \, \mathrm{mag^2}$, $(0.02 \pm 0.02) \, \mathrm{mag^2}$, and $(0.07 \pm 0.04) \, \mathrm{mag^2}$ for the J, H, and K bands, respectively. 
While these results indicate that we cannot draw definitive conclusions about the intrinsic variability of J1544 in the NIR band, potentially related to the bimodality between high and low modes, visual inspection of the light curves in Fig.~\ref{Fig:lc_nics} and the slightly higher excess variance in the K band may suggest that further investigations are required to better characterize the variability pattern at these wavelengths. To our knowledge, these are the first NIR light curves reported for this source, and future high-time resolution NIR observations with simultaneous X-ray coverage could provide valuable insights.

\subsection{Radio variability}
J1544 appeared pointlike in all our radio observations, with an angular resolution ranging from 3.5\arcsec\, to 6\arcsec. The source was not detected during our VLA observation on 2024 February 8, with a $3\sigma$ flux density upper limit of $\sim$$8 \, \mathrm{\mu Jy}$ at $6 \, \mathrm{GHz}$. While we observed an excess of emission at the target's position, the low signal-to-noise ratio ($\sim$3) and the presence of nearby noise peaks with comparable flux densities prevent us from confirming a detection.
Considering the radio/X-ray anti-correlation observed for J1023 -- where the radio flux increases as the system switches from the high to the low X-ray mode -- \citep{Bogdanov2018, Baglio_CotiZelati_2023A&A} and the hint of a similar trend for J1544 \citep{Gusinskaia_2025MNRAS}, we performed a deeper analysis separately extracting radio images during X-ray high and low modes simultaneously observed with \textit{XMM-Newton}. Considering the total VLA on-source time of $\sim$2.89 hr (see Sect.~\ref{sec:VLA_analysis}) and that J1544 spent $\sim$58\% and $\sim$33\% of the \textit{XMM-Newton} observation time in the high and low modes, respectively (see Sect.~\ref{sec:X-ray_emission}), we estimated that the radio images extracted during simultaneous X-ray high modes correspond to $\sim$1.68~hr, while those extracted during low X-ray modes correspond to $\sim$0.95~hr. However, the source remained undetected during either mode, with a $3\sigma$ flux density upper limit of $13 \, \mathrm{\mu Jy}$ during the X-ray low modes and $11 \, \mathrm{\mu Jy}$ during the high modes.
Notably, during the X-ray low modes, we observed an excess emission with a signal-to-noise ratio of $\sim$4 (peak of $\sim$$17 \, \mathrm{\mu Jy/b}$). However, a definitive detection remains challenging due to the previously mentioned limitations, particularly the presence of noise peaks with comparable flux densities.
For comparison, \citet{Gusinskaia_2025MNRAS} reported a VLA flux density of $(28.6 \pm 2.1) \, \mathrm{\mu Jy}$ during a 2019 observation in the C-band ($4-8$~GHz), which increased to $(56.6 \pm 8.3) \, \mathrm{\mu Jy}$ during simultaneous X-ray low-mode intervals. However, the total on-source time during their VLA observation was $\sim$4.2 hr. Similarly to our findings, they also reported that J1544 spent $\sim$33\% of the time in the X-ray low mode, meaning that their radio images extracted in the low mode corresponded to a longer total exposure time than ours.

J1544 was not detected in either of the ATCA observations, with a $3\sigma$ upper limit of $27 \, \mathrm{\mu Jy}$ at $7.25 \, \mathrm{GHz}$ (combining the C and X bands) in the first epoch, and $21 \, \mathrm{\mu Jy}$ at $7.25 \, \mathrm{GHz}$ in the second epoch. For both observations, the RMS noise was higher than expected due to the presence of strong RFI in our spectral windows. Stacking together the two ATCA epochs yielded a $3\sigma$ flux density upper limit of $15 \, \mathrm{\mu Jy}$ at $7.25 \, \mathrm{GHz}$, which is slightly higher than the VLA one. 

\begin{figure*}[h]
   \centering
   \includegraphics[width=0.9\textwidth]{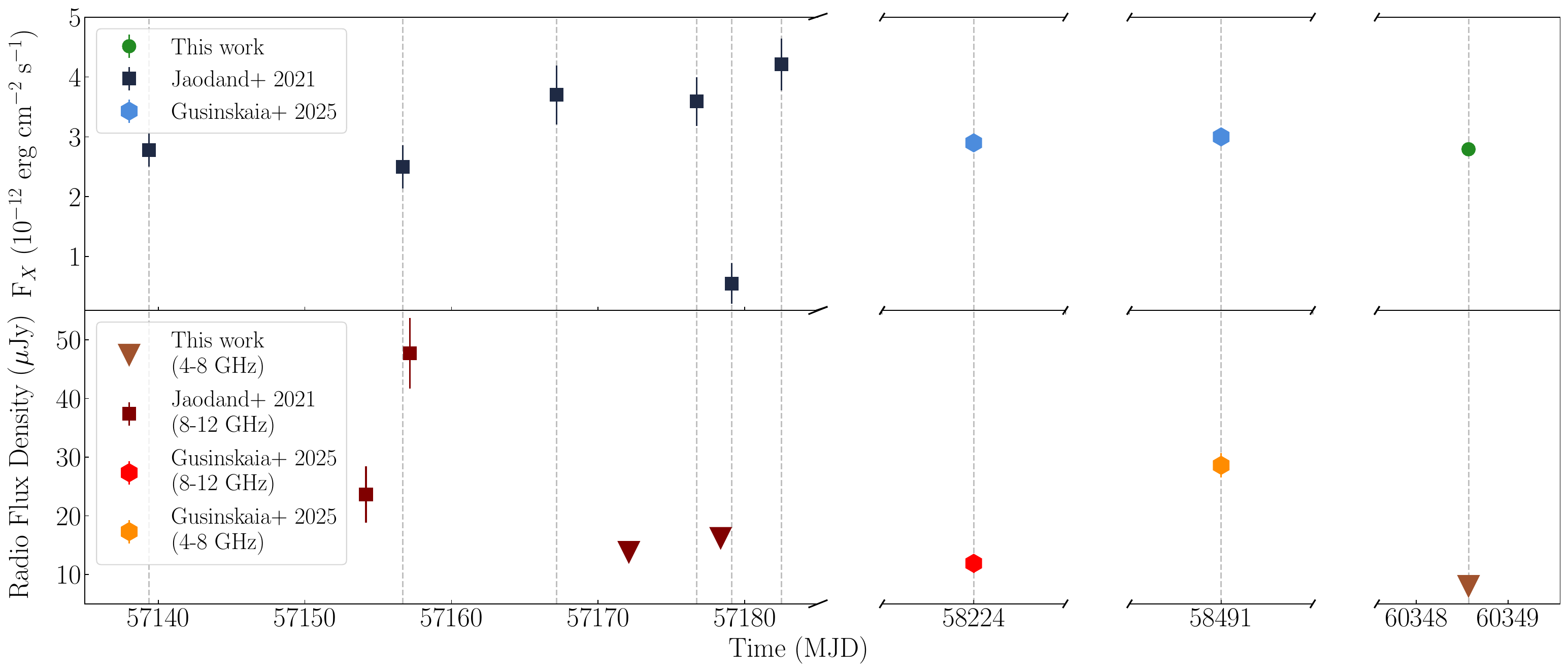}
      \caption{Radio variability of J1544 compared to its relatively stable X-ray flux over the years. This figure extends Fig.~1 from \citet{Jaodand_2021ApJ}, incorporating observations from \citet{Gusinskaia_2025MNRAS} and this work. \emph{Top panel}: $1-10$~keV unabsorbed X-ray flux (estimated using the entire observations, including both high and low modes) from \citet{Jaodand_2021ApJ} (dark blue squares), \citet{Gusinskaia_2025MNRAS} (light blue hexagons), and this work (green point). For consistency, we extracted the flux in the $1-10$~keV band from our \textit{XMM-Newton} observation, obtaining $F_X=(2.79 \pm 0.04) \times 10^{-12} \, \mathrm{erg \, cm^{-2} \, s^{-1}}$, following the procedure detailed in Sect.~\ref{sec:X-ray_emission}. The low X-ray flux around MJD 57179 is likely due to the source remaining in the X-ray low mode throughout this short (418 s) observation, as noted by \citet{Jaodand_2021ApJ}.
      \emph{Bottom panel}: VLA radio flux density in the X band ($8-12$~GHz) from \citet{Jaodand_2021ApJ} (dark red squares for detections and downward triangles for $3\sigma$ upper limits), and from \citet{Gusinskaia_2025MNRAS} at $8-12$~GHz (X band) around 58224~MJD (red hexagon), and at $4-8$~GHz (C band) around 58491~MJD (orange hexagon). The $3\sigma$ upper limit of $\sim$$9 \, \mathrm{\mu Jy}$ at $4-8$~GHz (C band) from this work is indicated by the brown downward triangle.
      For consistency with \citet{Jaodand_2021ApJ}, the plot shows the radio flux density from \citet{Gusinskaia_2025MNRAS} and the upper limit derived in this work both extracted using the entire VLA observation, rather than restricting to intervals simultaneous with X-ray low modes.
      The dashed vertical lines mark the epochs of X-ray observations for reference.}\label{Fig:radio_variability}
\end{figure*}

Figure~\ref{Fig:radio_variability} extends Fig.~1 from \citet{Jaodand_2021ApJ}, integrating results from \citet{Gusinskaia_2025MNRAS} and this work, to illustrate the radio variability of J1544 against a relatively stable X-ray flux over the years. 

\subsection{Spectral energy distribution} \label{sec:SED_J1544}
To derive the SED from the UV to the X-rays, we analyzed the \textit{HST} NUV spectrum between 2000 and 3000 \r{A} to avoid high background noise outside this range.
We corrected for interstellar extinction using an absorption column density of $\mathrm{N_H}=(1.52 \pm 0.06) \times 10^{21} \, \mathrm{cm^{-2}}$ as estimated from X-ray spectral fitting (see Sect.~\ref{sec:X-ray_emission}), and the empirical relation for the optical extinction $A_V=\mathrm{N_H}/[(2.87 \pm 0.12) \times 10^{21} \, \mathrm{cm^{-2}}]$ \citep{Foight_2016ApJ}.
Using the commonly adopted Milky Way average value for the total-to-selective extinction ratio, $R \equiv A_V/E(B-V) = 3.1$ \citep[e.g.,][]{Johnson_1965ApJ, Schultz_Wiemer_1975A&A, Whittet_Van_Breda_1980, Fitzpatrick_Massa_1999}, we determined a color excess of $E(B-V) = (0.17 \pm 0.1) \, \mathrm{mag}$.
To account for interstellar reddening and correct the flux values across different frequencies, we employed the \texttt{dust\_extinction} Python package with extinction curves from \citet{Gordon2024} \citep[see also][]{Gordon_2009, Fitzpatrick_2019, Gordon_2021, Decleir_2022}.

\begin{figure*}
   \centering
   \includegraphics[width=0.58\textwidth]{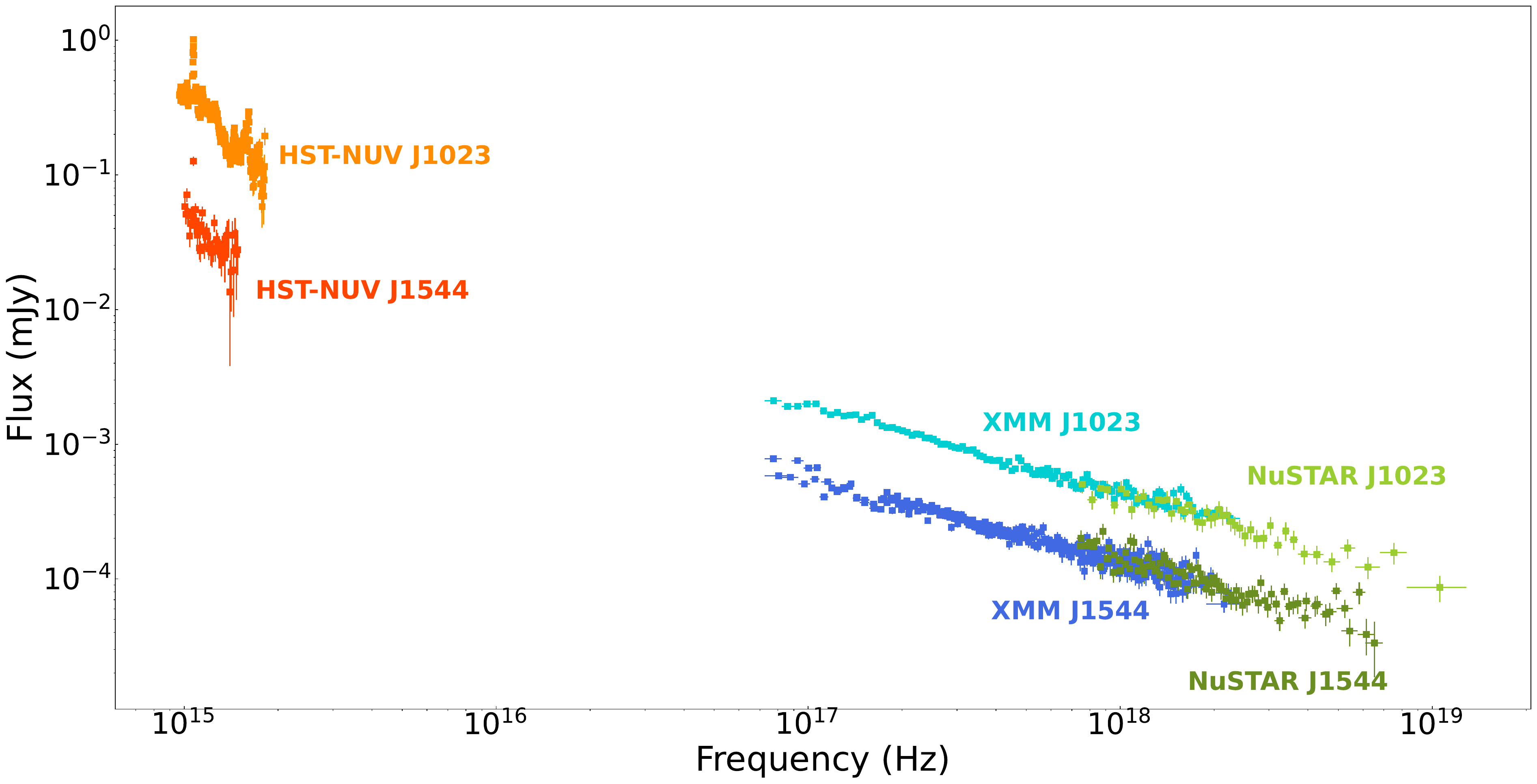}
   \includegraphics[width=0.38\textwidth]{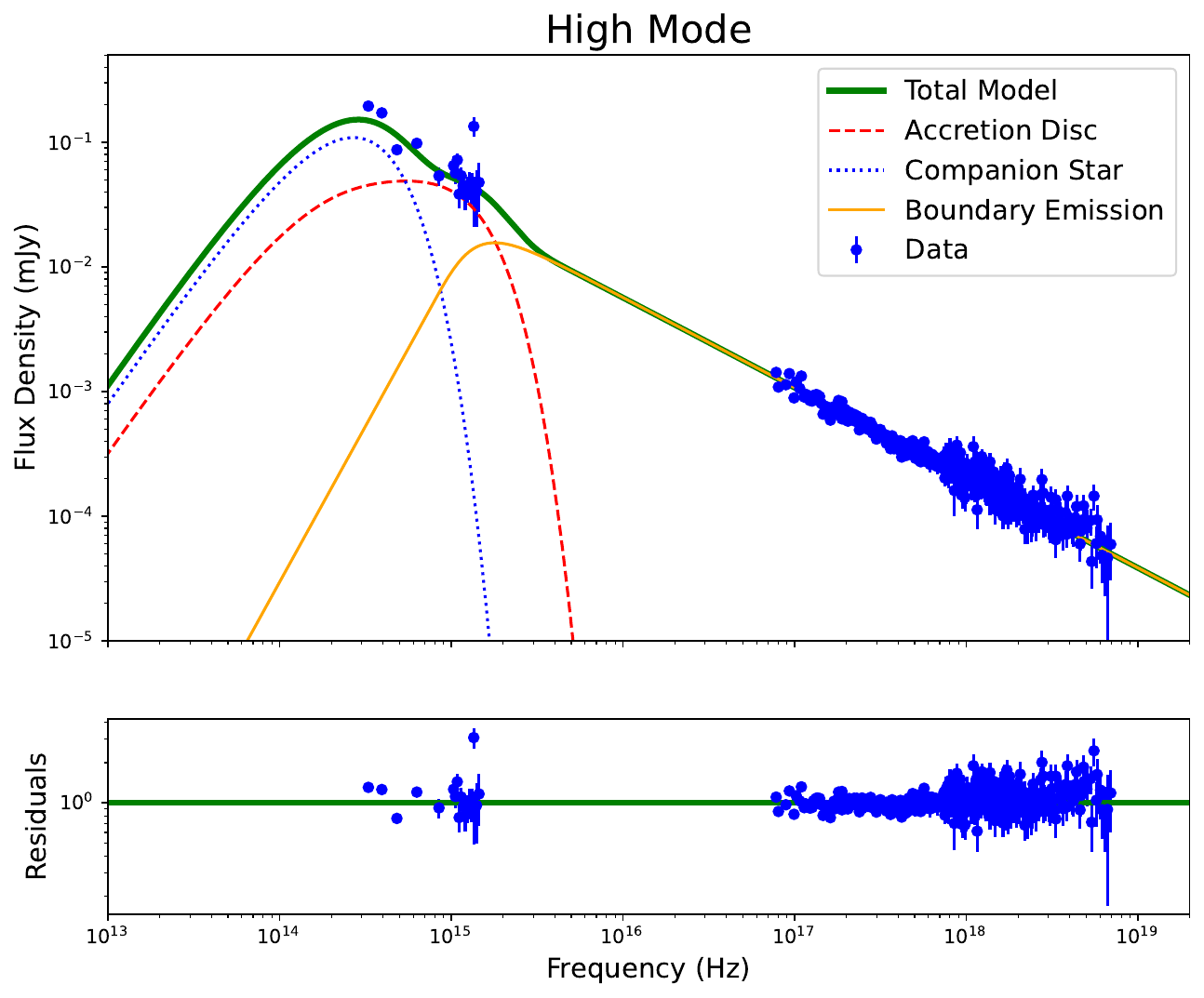}
      \caption{Unabsorbed broadband SED of J1544. \textit{Left panel}: The SED of J1544 from UV to X-rays, compared with that of J1023 from \cite{Miraval_Zanon_2022}. For J1023 and J1544, \textit{HST} data are shown in orange and red, \textit{XMM-Newton} data in light blue and dark blue, and \textit{NuSTAR} data in light green and dark green, respectively. The \textit{HST} spectrum is plotted from 165 to 310 nm (i.e., $\sim$$(1.0-1.8) \times 10^{15} \, \mathrm{Hz}$) for J1023 and from 200 to 300 nm (i.e., $\sim$$(1.0-1.5) \times 10^{15} \, \mathrm{Hz}$) for J1544. The UV data were rebinned using the \texttt{coronagraph} Python package \citep{Robinson_2016, Lustig-Yaeger_2019JOSSF} with a low-resolution wavelength grid of width 4 for J1023 and width 20 for J1544. \textit{Right panel}: The SED of J1544 from the optical band to X-rays, extracted during X-ray high modes observed with \textit{XMM-Newton}, along with the best-fitting model (see text for details). Due to the lack of simultaneous X-ray observations, we include all available optical data from GTC/HiPERCAM. The red dashed, blue dotted, and orange solid lines represent the contributions from the accretion disk, companion star, and boundary region, respectively, while the green solid line shows their combined emission. The ratio between the data points and the best-fitting model is shown in the bottom panel.}\label{Fig:SED}
\end{figure*}

The left panel of Fig.~\ref{Fig:SED} shows the SED of the total emission from UV to X-rays for J1544, compared to J1023 as reported by \citet{Miraval_Zanon_2022}. The striking similarity strongly supports the classification of J1544 as a very promising candidate tMSP in the sub-luminous disk state.

We then extracted the SED only during the high modes (right panel of Fig.~\ref{Fig:SED}), where the source spends most of its time, ensuring better statistical quality compared to the low modes. This choice is further justified by the absence of a clear bimodal behavior in the UV light curve as well as by the challenges in modeling the SED of J1544 due to its poorly constrained parameters.
We fitted the SED adopting a model based on the mini-pulsar nebula scenario (see Sect.~\ref{sec:intro} and \citealt{Baglio_CotiZelati_2023A&A} for a detailed discussion). In this framework, the X-ray emission primarily originates from synchrotron radiation produced at the boundary region between the pulsar wind and the inner accretion flow. In the SED extracted during the high modes of J1023 and analyzed by \citet{Baglio_CotiZelati_2023A&A}, an additional contribution to the X-ray emission was attributed to an optically thin synchrotron component from the base of a compact jet. 
However, in our analysis of J1544, we did not include this component as the available data do not provide sufficient constraints to model it reliably. Since this component contributed only $\sim$4\% of the X-ray emission, its omission does not significantly impact our results.
At lower frequencies, the model includes blackbody emission from the irradiated companion star and a multi-color blackbody component from the accretion disk. Following similar reasoning as for the optically thin synchrotron component, we did not include the optically thick synchrotron emission from the compact jet, which was considered for J1023 by \citet{Baglio_CotiZelati_2023A&A}.
To better constrain the contributions from the accretion disk and companion star, we included all available optical data from GTC/HiPERCAM, as no strictly simultaneous X-ray and optical observations were available.

\begin{figure*}
   \centering
   \includegraphics[width=0.73\textwidth]{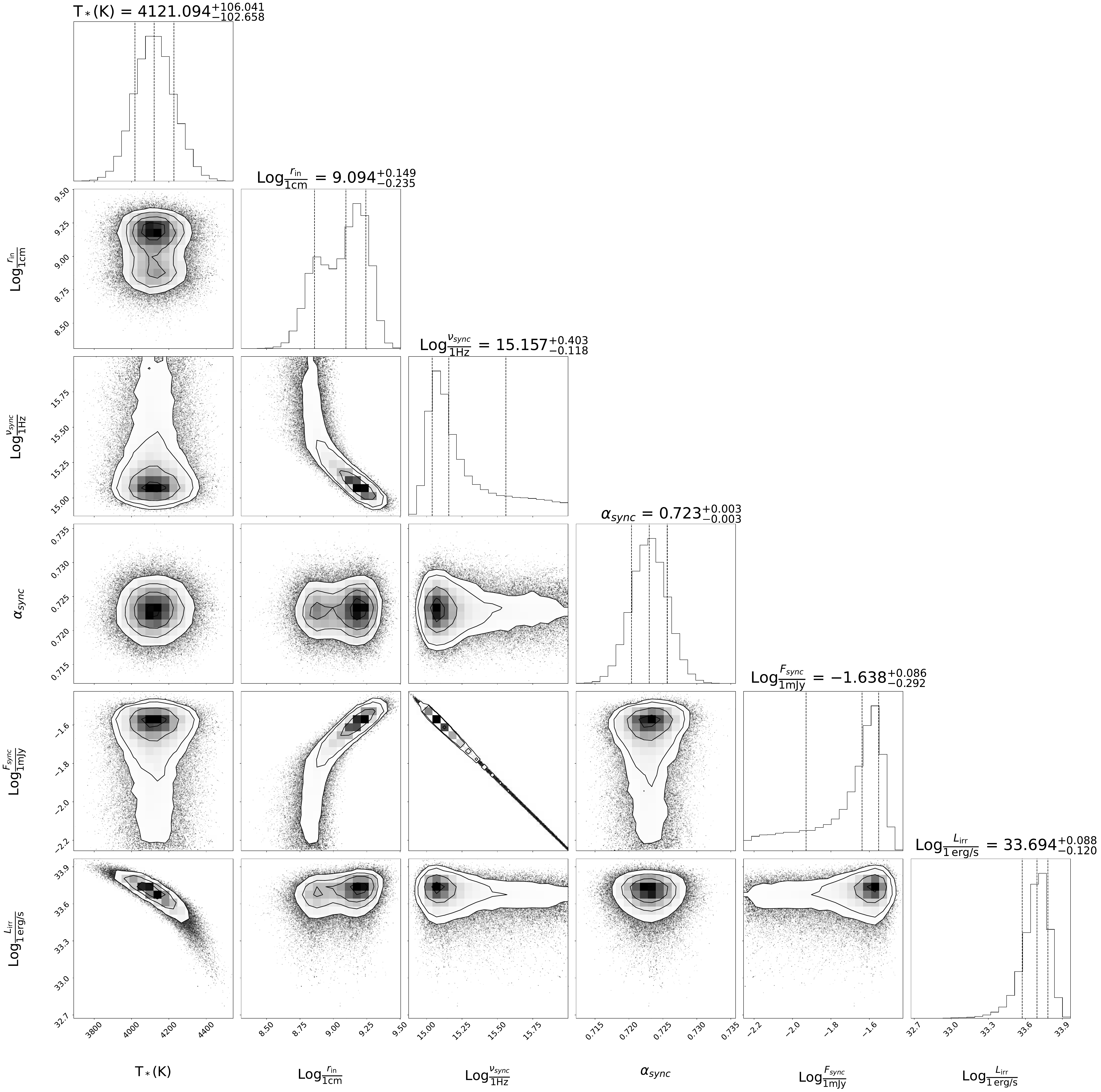}
   \caption{Corner plot displaying the posterior probability distributions of the parameters obtained from the MCMC sampling algorithm. The diagonal panels show the probability density of each parameter, with solid lines representing the distribution and vertical dashed lines marking the 16th, 50th, and 84th percentiles. The median values and corresponding 1$\sigma$ uncertainties are indicated above each panel. The off-diagonal panels illustrate the 2D posterior distributions, with contour lines representing the 1$\sigma$, 2$\sigma$, and 3$\sigma$ equivalent bounds.} \label{fig:corner_plot}
\end{figure*}

The free parameters in the model are the surface temperature of the companion star ($T_{*}$), the irradiation luminosity ($L_{\rm irr}$), the inner radius of the accretion disk where optical/UV emission becomes relevant ($r_{\rm in, opt/UV}$), the peak frequency of the synchrotron emission ($\nu_{\rm sync}$), the slope of the optically thin synchrotron emission component ($\alpha_{\rm sync}$), and the normalization of the synchrotron emission component ($F_{\rm sync}$). The fixed parameters in the model include the source’s distance ($D=3.8 \, \mathrm{kpc}$; this value was estimated by \citealt{Britt_2017ApJ}, but recent parallax measurements of the proposed optical counterpart by \citealt{Koljonen_Linares_2023MNRAS} suggest it should now be considered as an upper limit), the companion star radius ($R_{\rm c}= 0.75 \, \mathrm{R_\odot}$, assuming the upper limit on the companion mass of $M_{\rm C}= 0.7 \, \mathrm{M_{\odot}}$ from \citealt{Britt_2017ApJ} and the main-sequence mass-radius relationship), the albedo of the companion star ($\eta_{*} = 0.1$), and the binary separation ($a$). The latter is given by $a=[G(M_{\rm NS}+M_{\rm C})P_{\rm orb}^2/(4\pi)^2]^{1/3}$, where $M_{\rm NS}= 1.4 \, \mathrm{M_{\odot}}$ is assumed for the NS mass, $P_{\rm orb}=5.8$\,hrs is the binary orbital period \citep{Britt_2017ApJ}, and $G$ is the gravitational constant. We also fixed the X-ray albedo of the disc to 0.95 \citep{Chakrabarty_1998ApJ} and the mass transfer rate to $10^{-10}\, \mathrm{M_{\odot} \, yr^{-1}}$ (\citealt{Verbunt_1993ARA&A}; see \citealt{Baglio_CotiZelati_2023A&A} for further details). 
We performed a Markov Chain Monte Carlo (MCMC) sampling to investigate the posterior probability distribution of the model's parameter space. The best-fit values for the free parameters are provided in Table~\ref{Tab:fit_res}, where they are also compared with the results from the modeling of the SED of J1023 in the high mode, as reported by \citet{Baglio_CotiZelati_2023A&A}. Each parameter is determined as the median of its marginal posterior distribution, with the $1\sigma$ credible intervals derived from the 16th to 84th percentiles of the posterior samples. The selected prior distributions are provided in the last column of Table~\ref{Tab:fit_res}. Figure~\ref{fig:corner_plot} displays the corresponding histograms of the sampled parameters.

\begin{table*}[]
    \centering
    \caption{Results of the model fit of the SED in the high mode for J1544 (this work) and J1023 \citep{Baglio_CotiZelati_2023A&A}.}
    \resizebox{\textwidth}{!}{
    \begin{tabular}{cllrclr}
\hline \hline
Component             		& Parameter 					& \multicolumn2c{Posterior} 	& 					& \multicolumn2c{Prior bounds}\\ \cline{3-4} \cline{6-7} 
\noalign{\vskip 0.15cm}
                 			&                  					& J1544   					& J1023$^{(\ast)}$  		&&  J1544   						&   J1023$^{(\ast)}$\\ \hline 
Irradiated star   			& $T_{*}$ (K) 					& $4115^{+103}_{-98}$ 		& $6141^{+33}_{-30}$ 	&& $\mathcal{N}(4100, 100)^{(\star)}$ 	&  $\mathcal{N}(6128, 33)$\\   \vspace{0.3cm}
                  			& Log [$L_{\rm irr}$ / 1~erg/s] 		& $33.7 \pm 0.1$ 			& $33.8^{(\dagger)}$ 	&& $\mathcal{U} (30,40)$ 				& --\\ \vspace{0.35cm}
Accretion disc    		& Log [$r_{\rm in,opt/UV}$ / 1~cm] 	& $9.09 \pm 0.2$ 			& $9.49\pm 0.02$ 		&& $\mathcal{U}(7, 12)$ 				& $\mathcal{U}(7, 12)$\\   \vspace{0.05cm}
Boundary emission  		& Log [$\nu_{\rm sync}$ / 1~Hz] 	& $15.2^{+0.4}_{-0.1}$ 		& $14.3^{+0.4}_{-0.3}$ 	&& $\mathcal{U}(10, 16)$ 				& $\mathcal{U}(12, 16)$\\ \vspace{0.05cm}
                        			& $\alpha_{\rm sync}$ 			& $-0.723\pm 0.003$ 		& $-0.61\pm 0.01$ 		&& $\mathcal{U}(0.4, 1.0)$ 			& $\mathcal{U}(-0.9, -0.4)$ \\ \vspace{0.35cm}
                        			& Log [$F_{\rm sync}$ / 1~mJy] 	& $-1.6^{+0.1}_{-0.3}$ 		& $-1.1_{-0.3}^{+0.2}$ 	&& $\mathcal{U}(-5, 5)$ 				& $\mathcal{U}(-3, 2)$ \\  \vspace{0.05cm}
Compact jet       		& Log [$\nu_{\rm break}$ / 1~Hz] 	& -- 						& $13.4\pm 0.3$ 		&& -- 							& $\mathcal{U} (10,14)$\\ \vspace{0.05cm}
                				& $\alpha_1$					& -- 						& $0.18\pm 0.06$ 		&& -- 							& $\mathcal{N}(0.2,0.2)$\\ \vspace{0.05cm}
                				& $\alpha_2$ 					& -- 						& $-0.78\pm 0.04$ 		&& -- 							& $\mathcal{N}(-0.7, 0.2)$\\ \vspace{0.35cm}
                				& Log [$F_0$ / 1~mJy] 			& -- 						& $-0.6\pm0.2$ 		&& -- 							& $\mathcal{U}(-2, 0)$\\   
\hline
\end{tabular}
}
\label{Tab:fit_res}\\
\parbox{1.0\textwidth}{
{\bf Notes.} We report the median values of the posterior distributions for the model parameters, along with their lower and upper uncertainties, for J1023 (from \citet{Baglio_CotiZelati_2023A&A} and indicated as $^{(\ast)}$, including prior assumptions from references therein) and J1544 (this work). These uncertainties correspond to the 15.9th and 84.1st percentiles of the posterior distributions for each parameter. The final column provides the prior bounds applied to all parameters, where $\mathcal{N}(\mu, \sigma)$ represents a normal distribution with a mean $\mu$ and variance $\sigma^2$, and $\mathcal{U}(a,b)$ denotes a uniform distribution in the $(a,b)$ interval. 
$^{(\dagger)}$ Fixed at $6.5 \times 10^{33} \, \mathrm{erg/s}$ \citep{Baglio_CotiZelati_2023A&A}.
$^{(\star)}$ From \citet{Britt_2017ApJ}; see Sect~\ref{sec:discussion} for a detailed discussion on the irradiation component.}
\end{table*}

\section{Discussion} \label{sec:discussion}
We have presented the most extensive high-time resolution multi-wavelength campaign on the candidate tMSP J1544. 
The observed bimodality in X-ray intensity between high and low modes, along with an X-ray spectrum well described by an absorbed power law with a photon index of $\sim$1.6 -- characteristic of tMSPs in the intermediate, sub-luminous disk state \citep[see][and references therein]{Papitto_deMartino_2022ASSL} -- further supports the classification of this source as a promising candidate.
We also aimed to investigate a bimodal variability pattern in the UV data through simultaneous \textit{XMM-Newton} and \textit{HST} observations. However, due to the faint nature of the source and the variability in the background, we could not draw definitive conclusions regarding the presence of high and low modes in the UV band, which were observed for the brighter prototype of tMSPs, J1023 \citep{Jaodand_2021ApJ, Miraval_Zanon_2022, Baglio_CotiZelati_2023A&A}. 

The GTC/HiPERCAM light curves show flickering and dipping activities that mirror the X-ray variability. 
Moreover, they provide clear evidence of potential low modes, during which the optical emission is significantly redder than in high modes. 
This finding aligns with the mini-pulsar nebula scenario proposed by \citet{Papitto_2019ApJ}, which explains optical, UV, and X-ray pulsations from J1023 as due to synchrotron emission in the boundary region where the striped pulsar wind interacts with the inner accretion disk \citep[see also][]{Veledina_2019}. During the switch from high to low mode, characterized by the simultaneous disappearance of pulsations across all three bands, the inner flow is ejected, resulting in a decline in the optical, UV, and X-ray fluxes. Subsequently, as the inner flow begins to replenish, the system switches back to the high mode (\citealt{Papitto_2019ApJ, Baglio_CotiZelati_2023A&A}; see also \citealt{Bogdanov2018} for a similar interpretation). The reddening of the optical emission at low fluxes, observed both in this work and for the candidate tMSP CXOU~J110926.4$-$650224 in the sub-luminous disk state \citep{CotiZelati2024}, is interpreted as a residual, fainter optical emission originating from the cooler, outer regions of the disk during the low modes. This result -- along with other evidence such as the similar pulse shapes observed in the optical and X-ray bands, the simultaneous detection of pulsations occurring only during the high modes, and the pulsed SED described by a single power-law relationship \citep{Papitto_2019ApJ} -- indicates that during the high modes, the majority of the optical emission originates from the boundary region between the pulsar wind and the accretion disk, just like the X-rays.

While analyzing archival \textit{XMM-Newton}/OM data acquired in 2018, we identified hints of an optical flare around 58147.4 MJD with no corresponding feature in the X-ray light curve, marking the first such detection from this source. Specifically, after an observation gap around 58147.1~MJD, we observe a rise in optical flux followed by the flare-like feature. Similar optical spikes have been observed from J1023 with \textit{XMM-Newton}/OM (although they always occur in correspondence with an X-ray flare; see, e.g., the bottom panel of Fig.~7(c) from \citealt{Jaodand_2016}) as well as with other instruments \citep[e.g.,][]{Shahbaz_2015}.
A study of an 80-day-long uninterrupted Kepler monitoring campaign of J1023 in 2017 by \citet{Papitto_2018} revealed that optical flares occurred for $\sim$15.6\% of the time, a noticeably higher fraction than previously reported based on X-ray and optical observations. \citet{Kennedy_2018} later reported an even higher occurrence rate of $\sim$22\%, likely due to differences in the definition of a flare. Both studies concluded that the orbital dependence of these flares remains uncertain and is strongly influenced by the way flares are classified. However, \citet{Papitto_2018} found that optical flares were more frequently detected when the companion star was at superior conjunction in its orbit. If confirmed, this would suggest that at least some of the flares originate from the reprocessing of X-ray emission off the companion star’s surface. Given that the 2018 OM observations were conducted in white light (spanning from the UV to the redder V-band in the optical range), we speculate that the rise in the optical flux and the hint of a flare may originate from the outer regions of the accretion disk or the companion star. Future long-term optical monitoring with optical telescopes, ideally with as much simultaneous X-ray coverage as possible, will be necessary to identify additional optical flares from J1544, determine whether they have an X-ray counterpart, and ultimately shed light on the physical mechanisms driving these events.

To further probe the multi-band variability of J1544, we analyzed two NIR observations acquired in 2016 with TNG/NICS. The resulting light curves displayed significant variability, possibly mimicking the X-ray bimodality between high and low modes. However, the estimated values of excess variance indicate that we cannot definitively confirm intrinsic NIR variability in this source. Investigating NIR variability in tMSPs seems to be a challenging task. The NIR light curves of the confirmed tMSPs J1023 and XSS~J12270$-$4859 typically exhibit substantial variability across all filters in the sub-luminous disk state, including several flares and just a few, mildly pronounced dips that may correspond to the X-ray mode-switching \citep{deMartino_2010A&A, Saitou_2011PASJ, deMartino2014, Hakala_2017, Shahbaz_2018, Papitto_2019ApJ, Baglio2019}. In contrast, recent NIR observation of the candidate CXOU~J110926.4$-$650224 in the sub-luminous disk state revealed a prolonged 20-minute dip  \citep{CotiZelati2024}. Future NIR observations simultaneous with high-time resolution X-ray observations are needed to better characterize the NIR variability for various confirmed and candidate tMSPs and to explore potential connections with X-ray high and low modes. Such studies could indeed help constrain the origin of the NIR emission, and thus the expected variability pattern, by determining whether it arises from a jet \citep[e.g.,][]{Baglio2019}, the outer accretion disk, the companion star, or some combination of these components (see, e.g., \citealt{Papitto_2019ApJ} for tentative evidence of correlated NIR and X-ray variability for J1023; see also \citealt{Baglio_CotiZelati_2023A&A} for subsequent lack of flares at the low-to-high mode switch and of any clear signs of mode switching).

We performed radio observations with the VLA simultaneously with \textit{XMM-Newton} and later, in November and December 2024, with ATCA. In the following, we focus on the VLA observations, which provided more stringent upper limits because the ATCA observations were affected by higher-than-expected RMS noise due to the presence of significant RFI in our spectral windows. 
Even when VLA radio images were extracted separately during the X-ray high and low modes -- an analysis motivated by the anti-correlated variability observed between the radio and X-ray emissions in J1023 \citep{Bogdanov2018, Baglio_CotiZelati_2023A&A} and a similar trend suggested for J1544 \citep{Gusinskaia_2025MNRAS} -- we did not detect the radio counterpart to J1544 with VLA. Our non-detection supports the radio variability previously observed \citep{Jaodand_2021ApJ,Gusinskaia_2025MNRAS}, despite the relatively stable X-ray flux observed over time (see Fig.~\ref{Fig:radio_variability}). To quantify the X-ray flux stability, we used the estimates from \citet{Gusinskaia_2025MNRAS}, as they are, to our knowledge, the only ones in the literature derived exclusively during the high modes. Following the procedure outlined in Sect.~\ref{sec:X-ray_emission}, we extracted the 0.5–8 keV unabsorbed flux during the high mode and found $F_{\mathrm{unabs,H}} = (4.67 \pm 0.03) \times 10^{-12} \, \mathrm{erg \, cm^{-2} \, s^{-1}}$, showing stability within a factor of 1.5 compared to the \textit{Chandra} 2018 and 2019 observations reported by \citet{Gusinskaia_2025MNRAS}.

\citet{Jaodand_2021ApJ} observed J1544 with the VLA in the X-band ($8-12$~GHz) during four epochs in 2015. These observations were quasi-simultaneous with \textit{Swift}/XRT observations, which revealed unabsorbed $1-10$~keV X-ray flux values ranging from $(0.55 \pm 0.34) \times 10^{-12} \, \mathrm{erg \, cm^{-2} \, s^{-1}}$ to $(4.21 \pm 0.43) \times 10^{-12} \, \mathrm{erg \, cm^{-2} \, s^{-1}}$. During this period, the VLA radio flux density varied significantly, from $(47.7 \pm 6.0) \, \mathrm{\mu Jy}$ to $<13.8 \, \mathrm{\mu Jy}$ ($3\sigma$ upper limit).
This variability was further highlighted in subsequent VLA observations reported by \citet{Gusinskaia_2025MNRAS}. 
A 2018 X-band ($8-12$~GHz) observation yielded a flux density of $(11.9 \pm 1.6) \, \mathrm{\mu Jy}$, increasing up to $(14.2 \pm 2.9) \, \mathrm{\mu Jy}$ during simultaneous X-ray low-mode intervals. Similarly, a 2019 C-band ($4-8$~GHz) observation measured $(28.6 \pm 2.1) \, \mathrm{\mu Jy}$, increasing up to $(56.6 \pm 8.3) \, \mathrm{\mu Jy}$ during X-ray low-mode intervals. In both cases, the unabsorbed $1-10$~keV flux detected by \textit{Chandra} remained consistent with the range reported by \citet{Jaodand_2021ApJ}.
Interestingly, during our VLA observations in February 2024, conducted in the C-band, J1544 was not detected, even though the $1-10$~keV X-ray flux during our \textit{XMM-Newton} observation was $(2.80 \pm 0.04) \times 10^{-12} \, \mathrm{erg \, cm^{-2} \, s^{-1}}$, consistent with the 2015 \textit{Swift}/XRT values. In the bottom panel of Fig.~\ref{Fig:radio_variability}, we compare our stringent upper limit on the radio flux density with the 2019 measurement in the same band by \citet{Gusinskaia_2025MNRAS}. This comparison shows that if J1544 had maintained a similar flux density in our recent observation, we would have detected it. Our non-detection thus underscores the pronounced radio variability of the source. Radio variability without significant changes in X-ray flux or spectral state has been observed in other accreting NSs at higher accretion rates, underscoring the still-mysterious connection between accretion flow and the launching of outflows in these systems (see, e.g., \citealt{Marino_2023MNRAS, Panurach_2023ApJ, Pattie_2024ApJ}; see also recent evidence for compact jet formation in tMSPs by \citealt{Koljonen_2025MNRAS}). Future, longer radio observations of J1544 --ideally simultaneous with \textit{XMM-Newton} -- will be highly valuable not only for understanding its radio variability and potential correlation with X-ray high/low modes but also for providing new insights into the broader, poorly understood mechanisms that drive mass accretion and ejection in NS X-ray binaries.

A significant step toward confirming J1544 as a tMSP is the striking similarity of its broadband (from UV to X-rays) SED to that of J1023 as reported by \citet{Miraval_Zanon_2022} (see left panel of Fig.~\ref{Fig:SED}). The SED modeling in the high mode (see right panel of Fig.~\ref{Fig:SED}), along with the reddening of optical emission at lower fluxes observed with GTC/HiPERCAM and discussed above, supports the mini-pulsar nebula scenario. Table~\ref{Tab:fit_res} presents the best-fit parameter values for J1544, compared with those of J1023 from \citet{Baglio_CotiZelati_2023A&A}, further reinforcing the resemblance between the two sources. The main difference is that, while for J1023 the irradiation luminosity was fixed at the value of $L_{\mathrm{irr}}=6.5 \times 10^{33} \mathrm{erg/s}$ estimated by \citet{Shahbaz_2019MNRAS}, the limited knowledge of the system's parameters for J1544 required treating this parameter as a free variable. In the following, we discuss the obtained value of $L_{\mathrm{irr}}$. 

It is important to note that \citet{Britt_2017ApJ} found no significant evidence of phase-dependent temperature variations, suggesting minimal irradiation of the companion star. Therefore, we adopted the effective temperature of \citet{Britt_2017ApJ} as the prior bound on $T_{*}$ (see Table~\ref{Tab:fit_res}) to account for the emission of the non-irradiated companion star. They also noted that the absence of irradiation signatures in their light curves could be attributed to the system’s nearly face-on inclination. However, in our analysis, we find that the contribution of irradiation is necessary to model the broadband flux distribution, both in the star and the accretion disk components.
Keeping these considerations in mind, we now examine the relationship between the observed irradiation luminosity and the spin-down power.
Assuming isotropy, the energy flux from the pulsar wind at the location of the companion star is given by $F_{\mathrm{irr}} = \dot{E}/(4\pi a^2)$, where $\dot{E}$ is the spin-down power and $a$ is the orbital separation. The fraction of this energy intercepted by the companion star depends on its projected area, leading to $f = R_\mathrm{c}^2/(4a^2)$, where $R_\mathrm{c}$ is the companion star’s radius. Accounting for an efficiency factor $\eta$ that quantifies the reprocessing of the absorbed energy into observable radiation, the irradiation luminosity can be expressed as $L_{\mathrm{irr}} = \eta [R_\mathrm{c}^2/(4a^2)] \dot{E}$. Considering that $\eta \leq 1$ and the value reported in Sect.~\ref{sec:SED_J1544}, the previous relation implies that the spin-down power for J1544 should be $\dot{E} \leq (6 \pm 1) \times 10^{34} \, \mathrm{erg/s}$. While this approach may be valuable -- since it could, in principle, constrain the irradiation luminosity and, by extension, the spin-down power for confirmed and candidate tMSPs in the sub-luminous disk state that have not been observed in the rotation-powered state -- it should be interpreted with caution. Notably, the upper limit obtained for J1544 is an order of magnitude lower than the estimate derived using the empirical correlation between X-ray luminosity and spin-down power recently proposed by \citet{Xu2025}, combined with our estimation of the X-ray luminosity. In the future, should J1544 transition to the rotation-powered state, deep radio observations with single-dish telescopes could enable the detection of its rotational parameters and unambiguously provide a measure of its spin-down power.

\section{Conclusions} \label{sec:conclusions}
This paper presented the most extensive high-time resolution multi-wavelength campaign conducted on the candidate tMSP J1544 in the sub-luminous disk state. Our key findings are summarized as follows:
\begin{itemize}
    \item X-ray/UV emission: The X-ray data clearly exhibited a bimodal pattern between high and low modes. This is consistent with previous studies and strongly supports the classification of J1544 as a promising tMSP candidate in the sub-luminous disk state. The UV emission showed marked variability. In some cases, these variations appeared to correspond to changes in the X-ray intensity level. However, due to the source faintness and background variability, we were unable to definitively establish bimodal behavior in the UV data.
    \item Optical/NIR variability: High-time-resolution optical observations showed variability patterns similar to the X-ray modes, with notable reddening of optical emission at low fluxes, supporting the mini-pulsar nebula scenario. This suggests that during low modes, residual optical emission comes from the cooler, outer regions of the disk after the inner flow is ejected, while in high modes, most optical emission originates from the boundary region between the pulsar wind and the accretion disk, just like the X-rays. The NIR variability was too noisy to draw firm conclusions, emphasizing the need for further multi-wavelength monitoring.
    \item Optical flare candidate: We reported the first candidate optical flare in archival \textit{XMM-Newton}/OM data, which lacked an X-ray counterpart. This suggests the presence of complex variability mechanisms and motivates future coordinated optical and X-ray monitoring over extended periods.
    \item Radio flux variability: Although the X-ray flux of J1544 has remained relatively stable in recent years, the source exhibited significant radio variability, alternating between detections and non-detections under similar observational conditions. This underscores the complexity of mass ejection processes in tMSPs.
    \item SED modeling and spin-down power estimate: The broadband SED of J1544 was strikingly similar to that of the tMSP J1023 in the sub-luminous disk state, supporting the mini-pulsar nebula scenario. By treating the irradiation luminosity as a free parameter, we showed that irradiation is essential for modeling the broadband flux from both the companion star and the accretion disk. Using the relationship between irradiation luminosity and spin-down power, we estimated J1544's spin-down power to be an order of magnitude lower than predicted by the empirical correlation between X-ray luminosity and spin-down power for rotation-powered pulsars.
\end{itemize}

\begin{acknowledgements}
The research leading to these results has received funding from the European Union’s Horizon 2020 Programme under the AHEAD2020 project (grant agreement n. 871158).
This work is based on observations made with the Gran Telescopio Canarias (GTC), installed at the Spanish Observatorio del Roque de los Muchachos of the Instituto de Astrofísica de Canarias, on the island of La Palma (program ID: GTC129-23B). This work is based on data obtained with the instrument HiPERCAM, built by the Universities of Sheffield, Warwick, and Durham, the UK Astronomy Technology Centre, and the Instituto de Astrofísica de Canarias. Development of HiPERCAM was funded by the European Research Council, and its operations and enhancements by the Science and Technology Facilities Council.    
SiFAP2 and NICS observations were made with the Italian Telescopio Nazionale Galileo (TNG) operated on the island of La Palma by the Fundación Galileo Galilei of the INAF (Istituto Nazionale di Astrofisica) at the Spanish Observatorio del Roque de los Muchachos of the Instituto de Astrofisica de Canarias.
This publication makes use of data products from the Two Micron All Sky Survey, which is a joint project of the University of Massachusetts and the Infrared Processing and Analysis Center/California Institute of Technology, funded by the National Aeronautics and Space Administration and the National Science Foundation.
This work is also based on observations acquired with the NICER mission, a 0.2--12\,keV X-ray telescope operating on the International Space Station; the NuSTAR mission, a project led by the California Institute of Technology, managed by the Jet Propulsion Laboratory, and funded by NASA; and XMM-Newton, an ESA science mission with instruments and contributions directly funded by ESA
Member States and NASA. This research has made use of the NuSTAR Data Analysis Software (NuSTARDAS) jointly developed by the ASI Space Science Data Center (SSDC, Italy) and the California Institute of Technology (Caltech, USA). We also used software and tools provided by the High Energy Astrophysics Science Archive Research Center (HEASARC) Online Service.
The National Radio Astronomy Observatory is a facility of the National Science Foundation operated under cooperative agreement by Associated Universities, Inc. The Australia Telescope Compact Array is part of the Australia Telescope National Facility (grid.421683.a) which is funded by the Australian Government for operation as a National Facility managed by CSIRO. We acknowledge the Gomeroi people as the traditional owners of the Observatory site.\\

G.I. thanks the HST program coordinator, A. Vick  (STScI), for constant support in the observation planning and A. Fullerton  (STScI) for checking the scheduling processes. G.I. thanks S. Dieterich (STIS Team) for the support in the scientific data analysis.
G.I. and A.M.Z. also thank M. Cadelano for the useful discussion on HST data analysis.\\

G.I., A.P., F.C.Z., S.C., D.d.M., C.M., R.L.P. are supported by INAF (Research Grant ‘Uncovering the optical beat of the fastest magnetised neutron stars 620
(FANS)’) and the Italian Ministry of University and Research (MUR) (PRIN 2020, Grant 2020BRP57Z, ‘Gravitational and Electromagnetic-wave Sources in the Universe with current and next-generation detectors (GEMS)’). A.P. acknowledges support from the Fondazione Cariplo/Cassa Depositi e Prestiti, grant no. 2023-2560.
F.C.Z. acknowledges support from a Ramon y Cajal fellowship (grant agreement RYC2021-030888-I). N.R. and A.M. are supported by the European Research Council (ERC) under the European Union’s Horizon 2020 research and innovation programme (ERC Consolidator Grant “MAGNESIA” No. 817661). F.C.Z. and N.R. acknowledge support from grant SGR2021-01269 from the Catalan Government. This work was also supported by the Spanish program Unidad de Excelencia Marıa de Maeztu CEX2020-001058-M and by MCIU with funding from European Union NextGeneration EU (PRTR-C17.I1). T.D.R. is an INAF research fellow. D.F.T. has been supported by PID2021-124581OB-I00 funded by MCIU/AEI/10.13039/501100011033 and 2021SGR00426 as well as by funding from the European Union NextGeneration EU (PRTR-C17.I1) program.
\end{acknowledgements}

  \bibliographystyle{aa} 
  \bibliography{bib} 

\begin{appendix} 
\section{Spectral analysis of archival \textit{XMM-Newton} data} \label{Sec:appendix_spectral_analysis}
To enhance the statistics of the spectra separately extracted in high and low modes and to investigate significant changes in the photon index between these states, we analyzed \textit{XMM-Newton} archival observations of J1544 performed on 2014 February 16 (ObsID 0724080101; PI: Turriziani) and on 2018 January 28 (ObsID 0800280101; PI: Bogdanov).
ObsID 0724080101 is detailed in \cite{Bogdanov_2015ApJ}, whereas during ObsID 0800280101 the EPIC-pn operated in timing mode, and the two EPIC-MOS in small window mode. We followed the data analysis procedure described in Sect.~\ref{sec:XMM_data_analysis}, carefully filtering out the high background flaring activity observed in the 10–12 keV light curves for ObsID 0800280101 (see middle panel of Fig.~\ref{fig:OM_2018}).

\begin{figure}
    \centering
    \includegraphics[width=0.47\textwidth]{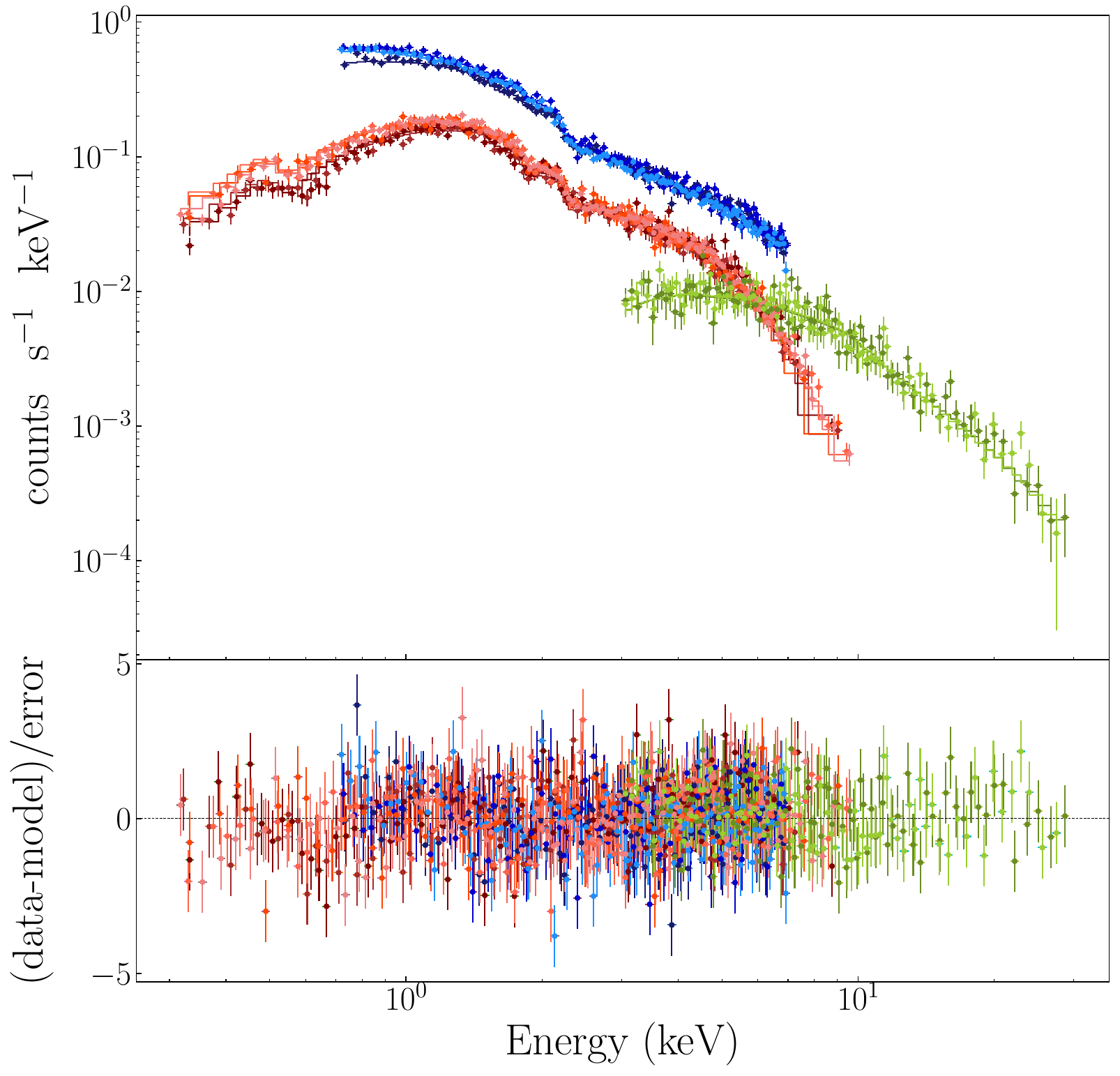}
    \includegraphics[width=0.47\textwidth]{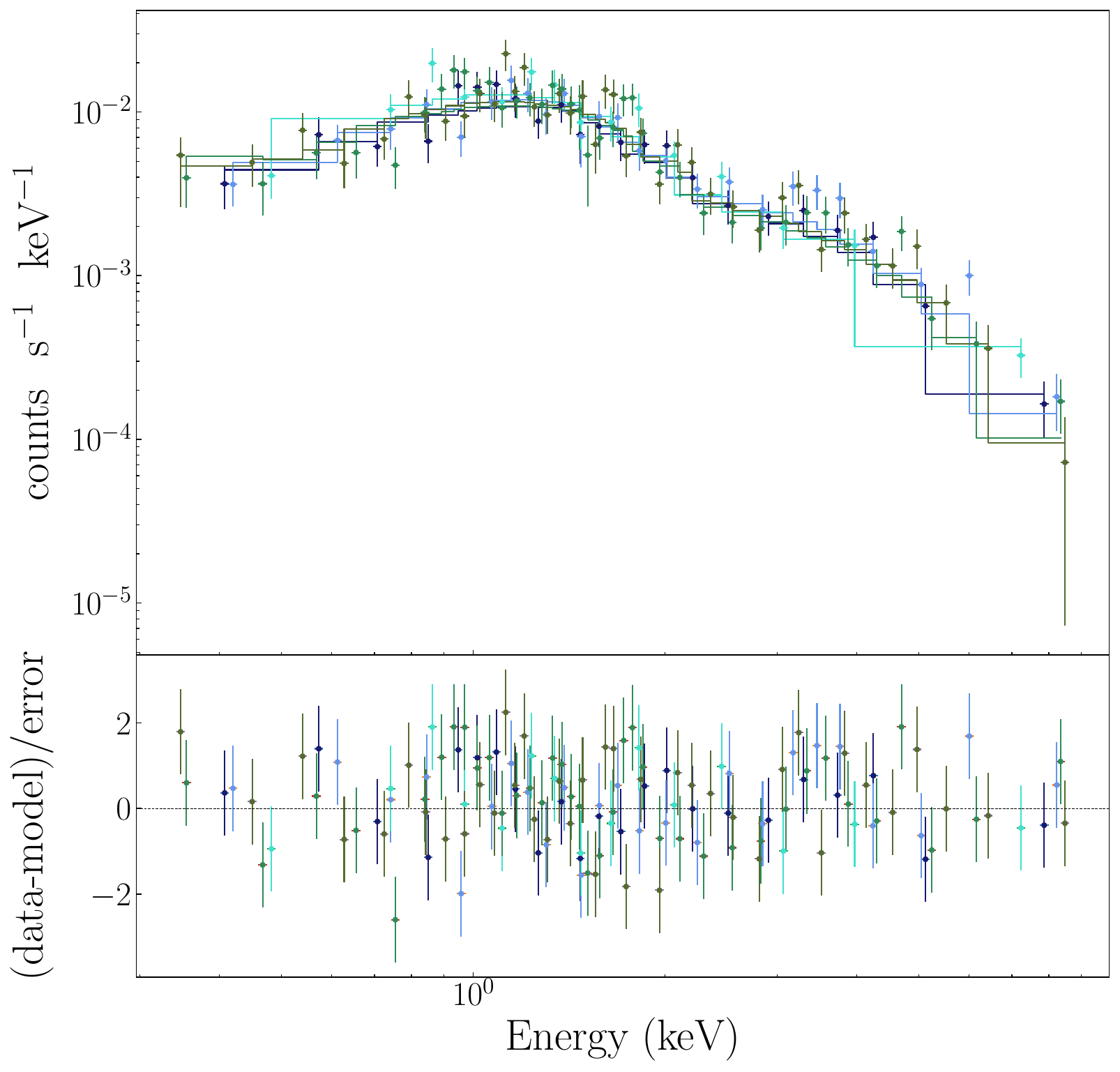}
    \caption{X-ray spectra extracted separately in the high-mode (\textit{top panel}) and low-mode (\textit{bottom panel}) intervals, along with the best-fitting models (see Appendix~\ref{Sec:appendix_spectral_analysis} for details). \textit{Top panel}: Various shades of blue denote the three EPIC-pn spectra, while different shades of red represent the spectra obtained with the two EPIC-MOS cameras. Light and dark green indicate \textit{NuSTAR} spectra. \textit{Bottom panel}: As discussed in the text, the low-mode spectrum includes only data for the two EPIC-MOS cameras.} \label{Fig:high_low_mode_spectrum_archival_obs}
\end{figure}

The high-mode spectrum (left panel of Fig.~\ref{Fig:high_low_mode_spectrum_archival_obs}) includes \textit{XMM-Newton} and \textit{NuSTAR} data as described in Sects.~\ref{sec:XMM_data_analysis} and \ref{sec:NuSTAR_data_analysis}, together with EPIC-pn and EPIC-MOS1 data from ObsID 0724080101, and EPIC-pn, EPIC-MOS1, and EPIC-MOS2 data from ObsID 0800280101. The low-mode spectrum (right panel of Fig.~\ref{Fig:high_low_mode_spectrum_archival_obs}) comprises EPIC-MOS1 data from ObsID 0724080101, and EPIC-MOS1 and MOS2 from ObsID 0800280101, excluding the background-dominated EPIC-pn data and low-statistic data from \textit{NuSTAR}.
The resulting total exposure is $\sim$467.97 ks during the high-mode intervals and $\sim$112.38 ks during the low-mode intervals.

We fitted the spectra with an absorbed power law, keeping the absorption column density fixed to the value obtained from the average spectral fitting (Sect.~\ref{sec:X-ray_emission}). 
We included a renormalization factor in the model to account for cross-calibration uncertainties between the two X-ray telescopes and different observations. For the high mode, these values remained consistent within 10\%, while for the low mode, they showed a larger discrepancy up to 14\%, likely due to the limited statistics in this mode.
For the high mode, we obtained $\Gamma_{\rm H} = 1.627 \pm 0.006$
and an unabsorbed 0.3-10\,keV flux $F_{\mathrm{unabs, H}} = (9.4 \pm 0.1) \times 10^{-12} \, \mathrm{erg \, cm^{-2} \, s^{-1}}$ 
($\chi^2$/d.o.f.=1100.12/1100), while for the low mode $\Gamma_{\rm L} = 1.66\pm0.06$ and $F_{\mathrm{unabs, L}} =
(0.38 \pm 0.03) \times 10^{-12} \, \mathrm{erg \, cm^{-2} \, s^{-1}}$ ($\chi^2$/d.o.f =139.0/136). 
No significant change in the photon index between high and low modes was reported by \citet{Bogdanov_2015ApJ}, and our results likewise show consistency within 1$\sigma$ across modes.

\clearpage
\section{Inspection of the optical flare candidate} \label{sec:appendix_flare}
\begin{figure*}
   \centering
   \includegraphics[width=0.8\textwidth]{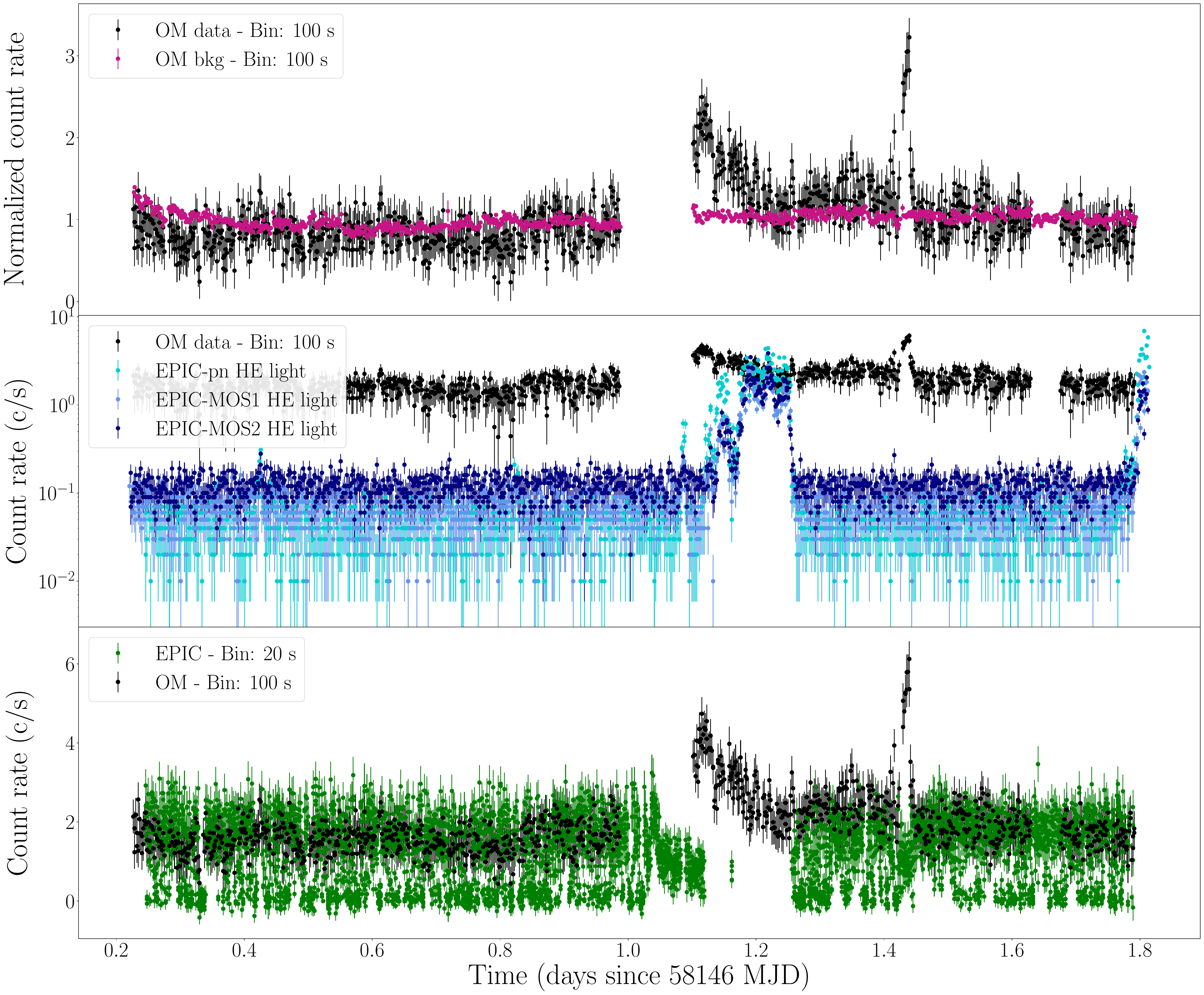}
   \caption{Temporal evolution of the optical emission as observed with \textit{XMM-Newton}/OM in 2018. \emph{Top panel}: OM light curve (black) and background light curve (purple), both binned at 100 s.
   \emph{Middle panel}: Comparison between the \textit{XMM-Newton}/OM light curve (100-s bins, shown in black) and the EPIC high-energy light curves in the 10-12 keV band (100-s bins; light blue for EPIC-pn, blue for EPIC-MOS1, and dark blue for EPIC-MOS2) extracted from the event files to identify intervals of flaring particle background. The y-axis is shown on a logarithmic scale for visual purposes.
   \emph{Bottom panel}: Overlap of the \textit{XMM-Newton}/OM light curve (100-s bins, shown in black) and the \textit{XMM-Newton}/EPIC background-subtracted light curve (20-s bins, shown in green).} \label{fig:OM_2018}
\end{figure*}

To investigate the nature of the flare-like event detected in the archival \textit{XMM-Newton}/OM data from 2018 January 28 (ObsID 0800280101; see Sect.~\ref{sec:hint_optical_flare}), we first confirmed that the observed feature was absent in the OM background light curve. During the OM processing using \texttt{omfchain}, the background level is derived from imaging-mode data using the recommended setting \texttt{bkgfromimage=yes}, rather than from the fast-mode window. This approach uses the average background obtained during the whole observing window as image mode and thus is assumed constant in the standard processing. For this reason, we instead used the background light curve as extracted in a region within the fast window  (see top panel of Fig.~\ref{fig:OM_2018}).
Since the flare-like feature is only present in the data and not in the background light curve, this supports its genuine nature.
We then carefully examined the \textit{XMM-Newton}/EPIC data for any intervals of flaring particle background. We used the \texttt{evselect} command with the following selection expressions: ``\#XMMEA\_EM \&\& (PI>10000) \&\& (PATTERN==0)'' for EPIC-MOS, and  ``\#XMMEA\_EP \&\& (PI>10000\&\&PI<12000) \&\& (PATTERN==0) for EPIC-pn'', using 100-s bins. As shown in the middle panel of Fig.~\ref{fig:OM_2018}, the optical feature does not coincide with periods of enhanced particle background in the X-rays. 
Finally, we visually inspected all individual OM images and verified that the source consistently remained within the Field of View throughout the observation. 

\end{appendix}

\end{document}